\input harvmac
\input epsf
\vskip 1cm

\noblackbox

 \Title{ \vbox{\baselineskip12pt\hbox{  Brown Het-1260 }}}
 {\vbox{
\centerline{  On  spherical harmonics for fuzzy spheres }
\centerline{ in diverse dimensions.     } }}

\centerline{$\quad$  { Sanjaye Ramgoolam }}
\smallskip
\centerline{{\sl Department of Physics}}
\centerline{{\sl Brown  University}}
\centerline{{\sl Providence, RI 02912 }}
\smallskip
\centerline{{\tt  ramgosk@het.brown.edu}}

 \vskip .3in

 We construct spherical harmonics for fuzzy spheres of even and odd
 dimensions, generalizing the correspondence between 
 finite matrix algebras and fuzzy two-spheres. 
 The finite  matrix algebras associated with 
 the various fuzzy spheres have a natural basis which falls  in correspondence
 with   tensor constructions  of irreducible representations 
 of orthogonal groups $SO(n)$. This basis is useful in describing 
  fluctuations around  various D-brane constructions  of fuzzy 
 spherical objects. The higher fuzzy spheres are non-associative
 algebras that 
 appear as  projections of associative
 algebras related to Matrices. The non-associativity ( as well 
 as the non-commutativity ) disappears in the leading large $N$ limit, 
 ensuring the correct classical limit.  
 Some simple aspects of the combinatorics of the fuzzy four-sphere 
 can be accounted by a heuristic picture of giant fractional
 instantons.

\def\Ga{ \Gamma } 
\def\cR{ {\cal{R}} } 
\def\cRp{ { { \cal{R}}_+ }  } 
\def\cRm{ { {\cal{R}}_-}  }
\def\Prp{ { {\cal{P}}_{\cal{R_+}} }  } 
\def\Prm{ { \cal{P}_{{\cal{R_-}} }  } }
\def\haf{ {1\over 2 } }

\def\Pr{ { {\cal{P}}_{\cal{R}} }  } 
\def\cAn{ {\cal{A}}_n }   
\def\chAn{ \hat {\cal{ A }}_n }
\def\hAg{   {\hat { \cal A} }^{(g)}_n  }

\lref\gkp{ H. Grosse, C. Klimcik and P. Presnajder,
``On finite  4-D Quantum Field Theory in noncommutative geometry,''
hep-th/9602115,  Commun.Math.Phys.180:429-438,1996. } 
\lref\dougf{ Free fermions and group theory }  
\lref\clt{ J.Castelino, S. Lee and W. Taylor IV, ``Longitudinal Five-Branes
as Four Spheres in Matrix Theory,'' Nucl.Phys.{\bf B526} (1998) 334, 
hep-th/9712105.}
\lref\zr{ Z. Guralnik and S. Ramgoolam, 
``On the Polarization of Unstable D0-Branes
 into Non-Commutative Odd Spheres,''JHEP 0102 (2001) 032,
 hep-th/0101001 } 
\lref\madr{J. Madore, ``The fuzzy sphere,'' { \it
 Class. Quant. Grav. }  
{\bf 9}  (1992 ) 69-87.  }
\lref\curtis{ Curtis } 
\lref\whn{ B. de Wit, J. Hoppe, H. Nicolai,
 ``On the quantum mechanics of supermembranes,'' Nucl.Phys.B305:545,1988  } 
\lref\myers{R. Myers, ``Dielectric Branes,'' JHEP {\bf 9912} (1999) 022, 
    hep-th/9910053  } 
\lref\bfss{T. Banks, W. Fischler , S. Shenker , L. Susskind ``M Theory
as a Matrix Model: a Conjecture,''
Phys.Rev.{\bf D55}, (1997) 5112, hep-th/9610043.} 
\lref\grt{ O. Ganor, S. Ramgoolam, W. Taylor, IV, ``Branes Fluxes and Duality
in M(atrix) Theory,'' Nucl.Phys.{\bf B492} (1997) 191, hep-th/9611202.}
\lref\bss{ T. Banks, N. Seiberg, S. Shenker, ``Branes from Matrices,''
  Nucl.Phys.{\bf B490} (1997) 91-106, hep-th/9612157.}
\lref\antram{ A. Jevicki and S. Ramgoolam, `` Noncommutative Gravity from the 
ADS/CFT correspondence,'' JHEP {\bf 9904} (1999) 032, hep-th/9902059.}   
\lref\hrt{  P.M.Ho, S.Ramgoolam and R.Tatar, ``Quantum Space-times 
and Finite N
Effects in 4-D Superyang-mills Theories,'' Nucl.Phys.{\bf B573} (2000) 364,
hep-th/9907145.}
\lref\holi{ P.M.Ho and M.Li, ``Fuzzy Spheres in ADS/CFT Correspondence
and Holography from Noncommutativity,'' hep-th/0004072.  } 
\lref\jmr{ A. Jevicki, M. Mihailescu, S. Ramgoolam, 
 ``Non-commutative spheres and the ADS/CFT correspondence,'' 
 JHEP 0010:008,2000, hep-th/0006239. } 
\lref\hidclass{ A. Jevicki, M. Mihailescu and  S. Ramgoolam,
``Hidden Classical Symmetry in Quantum Spaces at Roots of Unity: from Q
Sphere to Fuzzy Sphere, '' hep-th/0008186. } 
\lref\fro{ J. Frohlich, O. Grandjean and A. Recknagel, 
``Supersymmetric Quantum Theory and (Non-commutative) 
Differential Geometry,'' Commun.Math.Phys.{\bf 193} (1998) 527, 
hep-th/9612205    } 
\lref\ho{ P. M. Ho , ``Fuzzy Sphere from Matrix Model,'' 
JHEP {\bf 0012} (2000) 014,
hep-th/0010165. } 
\lref\mst{J. McGreevy, L. Susskind and N. Toumbas, 
``Invasion of the Giant Gravitons from Anti-de Sitter Space,''
JHEP {\bf 0006} (2000) 008, hep-th/0003075.}
\lref\fh{ W. Fulton and J. Harris, ``Representation Theory, ''
Springer-Verlag, 1991 } 
\lref\malstro{J. Maldacena and A. Strominger, 
``AdS(3) black holes and a stringy exclusion principle,'' 
JHEP 9812:005,1998,.  hep-th/9804085 } 
\lref\baracz{ A.O.Barut and R. Raczka, ``Theory of Group
Representations
 and Applications,'' World Scientific, 1976 }  
\lref\konsch{A. Konechny,  A. Schwarz,  ``Moduli spaces of maximally 
supersymmetric solutions on noncommutative tori and noncommutative
orbifolds, '' JHEP 0009:005, 2000;  hep-th/0005174  }
\lref\ztor{ Z. Guralnik and S. Ramgoolam, 
 ``From zero-branes to torons,''Nucl.Phys.B521:129-138,1998, 
   hep-th/9708089 }  
 \lref\napo{V. P. Nair and A. Polychronakos, 
``Quantum mechanics on the noncommutative plane and sphere,''
 Phys.Lett.B505:267-274,2001,  hep-th/0011172 } 
\lref\buk{ A. Buchel, 
`` Comments on fractional instantons in n=2 gauge theories, '' 
hep-th/0101056  } 
\lref\zk{ Z. Guralnik ``Strings and discrete fluxes of QCD,''
JHEP
0003:003,2000, hep-th/9903014 
  } 
\lref\bdie{ J. Brodie, 
`` Fractional branes, confinement, and dynamically generated
superpotentials'' hep-th/9803140 } 
\lref\mal{ J. Maldacena, ``Statistical entropy of near extremal
five-branes, '' Nucl.Phys.B477:168-174,1996  } 
\lref\fun{ Neil R. Constable, R. Myers, O.  Tafjord 
``Nonabelian brane intersections,'' hep-th/0102080  }
\lref\insmod{ C. Vafa and E. Witten, ``  A one loop test of string
duality,''Nucl.Phys.B447:261-270,1995 
, hep-th/9505053 }
\lref\corn{ L. Cornalba  and R. Schiappa, ``Nonassociative star
product deformations for d-brane worldvolumes in curved backgrounds, 
 hep-th/0101219  } 
\lref\pmnass{ P.M. Ho,  `` Making non-associative algebra associative ''
hep-th/0103024  }   
\lref\disc{ Discussions with N. Drukker, Z. Guralnik, P. Horava. } 
\lref\almy{ H. Awata, M.Li, D.Minic and T.Yoneya, 
``On the quantization of nambu brackets,'' hep-th/9906248; JHEP 0102; 013 } 
\lref\con{ A. Connes, ``Non-commutative geometry,'' Academic Press,
 1994.  }  
\lref\gdwa{ R. Goodman and N. Wallach, ``Representations and 
 Invariants of classical groups,'' 
 Cambridge University Press, 1998 } 
\lref\bv{ M. Berkooz and H. Verlinde, JHEP 9911:037,1999, hep-th/9907100   } 
\lref\fgr{ J. Frohlich, O. Grandjean, A. Recknagel,
 Commun.Math.Phys.203:119-184,1999 math-ph/9807006  } 
\lref\gomez{ C. Gomez, B.Janssen, P.J. Silva, hep-th/0011242 } 
\lref\mtv{ K. Millar, W. Taylor, M. van Raamsdonk, 
``D-particle polarizations with multipole moments 
 of higher dimensional branes,'' hepth/0007157 } 
\lref\bhp{ C. Bachas, J. Hoppe, B. Pioline, ``Nahm's equations,
$N=1^*$ domain walls, and D-strings in $ADS_5 \times S^5 $,''
hepth/0007067 } 
\lref\sah{V. Sahakian, ``On D0-brane polarization by tidal forces,''
hep-th/0102200 } 
\lref\intdi{Y. Gao, Z. Yang, ``Interactions between Dielectric
Branes,'' hep-th/0102151 }
\lref\kata{ D. Kabat and W. Taylor, Adv. Theor. Math. Phys. 2, 181,
1998, hepth/9711078 } 
\lref\sjr{ S.J. Rey,  hepth/9711081 } 
\lref\cmt{ N. Constable, R. Myers, O. Tafjord, Phys. Rev. D61, 106009 (2000), 
 hepth/9911136 } 
\lref\yol{ Y. Lozano, ``Non-commutative branes from M-theory,'' 
 hepth/0012137. } 
\lref\cvj{C.V. Johnson,  Int.J.Mod.Phys.A16:990,2001, hepth/0011008 }  
\lref\tv{ S. Trivedi and S. Vaidya, JHEP 0009 (2000) 041, hep-th/0007011 } 
\lref\postr{ J. Polchinski and M. Strassler, hepth/0003136 }  
\lref\horstro{ G. Horowitz, A. Strominger, `` Translations as inner
derivations 
 and associativity anomalies in open string field theory,''
Phys. Lett. B185, Vol. 2 } 
\lref\gms{ H. Grosse, J. Madore, H. Steinacker, 
``Field Theory on the q-deformed Fuzzy Sphere II: Quantization,'' 
   hep-th/0103164} 
\lref\kpt{ Dietmar Klemm, Silvia Penati, Laura Tamassia,
``Non(anti)commutative Superspace,''  hep-th/0104190 }


\Date{ April 2001 } 
\newsec{ Introduction. } 

 Non-commutative  spheres have found 
 a variety of physical applications.
 The non-commutative geometry \con\  of the fuzzy two-sphere 
 was first described in \madr. The fuzzy four-sphere 
 appeared in \gkp.   
 The fuzzy two-sphere was used in \whn\ in connection  
 with the quantization of the membrane. 
 The fuzzy 4-sphere was used \clt\  in the context of the Matrix Theory 
 of BFSS \bfss\  to describe time-dependent 
 4-brane solutions constructed from zero-brane degrees of freedom. 
 Non-commutative spheres were proposed 
 as  models of non-commutative space underlying the 
 stringy exclusion principle \malstro\ in \antram\ and 
 explored further in this context \hrt\holi\bv\jmr. 
 The fuzzy two-sphere described polarized D0-branes 
 in  a background three-form field strength in \myers. 
 Fuzzy spheres of diverse dimensions were found in Matrix theory 
 in \ho. 
 A general construction of odd  fuzzy spheres 
 was described in \zr\ and the fuzzy three sphere 
 was applied to the study of polarization of unstable 
 $D0$-branes.  Other studies of polarization-related effects 
 appear in \refs{ \mtv, \bhp, \intdi, \kata, \sjr, \cmt, \yol, \cvj,
 \postr }.

 In this paper we give a detailed 
 connection between various fuzzy spheres and Matrix Algebras. 
 This allows us to describe fuzzy spherical harmonics 
 which are related to a projection of Matrix algebras. 
 The complete $SO(m +1)$ decompositon of the Matrix algebras 
 is useful for describing fluctuations of Matrix 
 constructs of fuzzy spherical objects. 

 Section 2 reviews the fuzzy two-sphere. 
 Section 3 reviews some group theory of $SO(2k+1)$ 
 which is relevant in describing the fuzzy sphere 
 $S^{2k}$. Section 4 describes the Matrix Algebra 
 which is related to the fuzzy 4-sphere, and gives 
 its decomposition into representations of $SO(5)$.
 Section 4.2 explains that the algebra generated by 
 the coordinates of the fuzzy 4-sphere is isomorphic 
 to the Matrix Algebra.   
 Section 4.3 explains the projection, and the resulting
 non-associativity of the projected multiplication, 
  that is necessary 
 to get the algebra of functions on the 
 fuzzy 4-sphere $ \cAn(S^4)$, from the Matrix Algebra
 which we call $ \chAn ( S^4 )$.  Section 5 explains 
 the generalization to higher even spheres, giving 
 some details for the case of the fuzzy 6-sphere 
 and the 8-sphere. Section 6 reviews some group 
 theory of $SO(2k)$ which is necessary in the description 
 of the odd-dimensional  fuzzy sphere $ S^{2k-1}$. Section 7 decomposes 
 into $SO(4)$ representations the Matrix Algebra related 
 to the fuzzy three-sphere. It is useful to distiguish here, 
 $ \chAn( S^3)$ which is isomorphic to the Matrix algebra, 
 $ \cAn( S^3)$ which is the projected algebra of the fuzzy spherical 
 harmonics, and  $ \hAg( S^3)$ which is generated by the 
 coordinates of the fuzzy sphere under the matrix product.
  $ \hAg( S^3)$ is larger than  $ \cAn( S^3)$ but smaller 
 than  $ \chAn( S^3)$. Section 8 generalizes the discussion to higher 
 odd spheres, giving some explicit details for the five-sphere. 
 In section 9, we return to the non-associativity 
 mentioned in section 4.3 and prove that it vanishes in the 
 leading large $N$ limit. We expect this is a generic feature 
 for all the higher fuzzy spheres. Section 10 discusses 
 the geometric interpretation of the Matrix algebra related to the
 fuzzy 4-sphere in terms of the  physics on
 the geometrical 4-sphere obtained by projection. 
 We are lead to hints of fractional instantons exhibiting 
 behaviour reminiscent of the giant gravitons of \mst.

\newsec{ Review: The fuzzy 2-sphere } 

 The fuzzy 2-sphere \madr\  is defined as the algebra 
 generated by the three elements $S_3, S_+, S_-$ 
 obeying the relations of the $SU(2)$ Lie algebra : 
\eqn\surels{\eqalign{& [ S_{+}, S_{-} ] = 2S_3 \cr 
                     & [S_3, S_{+} ] = S_+ \cr 
                     & [ S_3,S_-] = -S_{-}. \cr }}
together with a constraint on the Casimir : 
\eqn\cas{ S_3^2 + { 1\over 2} (S_{+}S_{-}+   S_{-}S_{+} ) = J(J+1) } 
This algebra is infinite dimensional. For example 
 $S_-^l$, for any $l$,  are independent elements. 
 It admits, however, a finite dimensional quotient
 which is isomorphic to the algebra of $ N \times N$ 
 matrices where $N = n + 1 $with the definition $n=2J$.
 We will call this finite
 dimensional truncation $\chAn (S^2)$. As an algebra over the complex
 numbers 
this is  isomorphic  to  the algebra of $N\times N $ matrices
$ Mat_N(  C )$.

 It admits an action of the universal enveloping algebra 
 of $SU(2)$, by taking commutators. 
 Under this action of $U(SU(2))$, the $\chAn ( S^2) $
 decomposes as a direct sum of representations
 of integer spin $s$ with unit multiplicity with 
 $s$ ranging over integers $s$ from $1$ to $n= N-1  $.
\eqn\dec{ \chAn = \oplus_{ s=0}^{n} V_s } 

 Writing $S_1= {1 \over 2 } ( S_+ + S_- ) $ and $S_2 = {1\over 2 i}
 ( S_+ - S_- ) $,  
 representations of spin $s$ correspond to matrices of the form 
 $$f_{a_1,a_2, \cdots a_n } S^{a_1} S^{a_2} \cdots S^{a_n},  $$
 where the indices $a_1 \cdots a_n$ run from $1$ to $3$, 
 and $f$ is a traceless symmetric tensor.

\newsec{ Review: Some  group theory of $SO(2k +1)$ } 
 The following is a review of useful facts which can be found in 
 standard group theory books e.g \fh\gdwa\baracz. 
 Representations of $SO(2k+1)$ can be put in $1-1$ 
 correspondence with Young diagrams, 
 labelled by row lengths $ ( r_1, r_2, \cdots r_k )
 $, which obey the constraints $ r_1 \ge r_2 \ge ... \ge r_k
 $. We will often denote this vector of row lengths by 
 $ \vec r $. It also useful to describe the Young   
 diagrams by column lengths $ \vec c = ( c_1, c_2, \cdots )$. 
 The column lengths satisfy the restrictions 
 $ c_1 \ge c_2 \ge c_3 \cdots $ and $c_1 \le k $.
 Let $B$ be the number of boxes in the
 Young diagram, i.e  the sum of 
 row lengths or the sum of column lengths. 
 As far as classical $SO(2k + 1 )$ group theory is concerned 
 there is no cutoff on the number of columns. 
 In the application to fuzzy spheres, we will 
 often have a cutoff $n$ on the number  of columns, 
 or equivalently on the length of the first row.

 The Young diagram describes an irreducible 
 representation which arises from a subspace 
 of the vector space of tensor products of $B$ copies 
 of the  $(2k +1) $ dimensional fundamental   representation $F$. 
 Let $f_{\mu}$ be a set of basis vectors for this representation, 
 with $\mu$ running from $1$ to $2k+1$.   
A basis  vector in this tensor space is of the form 
 $ f_{\mu_1} \otimes f_{\mu_2} \otimes \cdots  \otimes f_{\mu_B}$, 
 and a general vector is a sum with coefficients 
 $ A ( \mu_1, \cdots , \mu_{B} )   $, 
 of the form 
 $$ \sum_{ \mu_1, \mu_2 \cdots, \mu_{B} }    A ( \mu_1, \cdots, \mu_{B} )~
 f_{\mu_1} \otimes f_{\mu_2} \otimes \cdots  \otimes f_{\mu_B}. $$

 On these tensors we can define  contraction 
 operations 
 $$ \sum_{\mu_i, \mu_j=1}^{n}  A ( \mu_1, \cdots, \mu_{B}  )$$
 which yield tensors of rank lower by two, 
  where $ (i,j)$ are any pair of distinct  integers  from $ 1$ to $B$.  
 The subset  of tensors with the property that 
 any contraction gives zero  are traceless tensors, 
 a generalization of traceless matrices. 

 The vectors of the  irreducible representation  are obtained by  
 applying, to the traceless tensors of 
 rank $B$,  a symmetrization procedure corresponding to 
 the Young diagram.
 This involves symmetrizing along the rows of the Young Diagram 
 and antisymmetrizing along the columns of the Young Diagram. 
 We can describe this more explicitly as follows.  
 Let us consider  $ A$ as a function of indices  $ \mu_{i}^{j} $,  labelled by 
 two parameters rather than one. This means that instead of writing 
 $A(\mu_1,\mu_2,\cdots \mu_B)$, we will write 
 $A(\mu^i_j)$. 
 Fixing the label $i$ corresponds to fixing a row, i.e $\mu^i_j$ 
 run over $j=1 \cdots r_i$ up to the length of the $i$'th row.
 For example $\mu^1_j$ has $j$ running over $1$ to $r_1$, 
 $\mu^2_j$ has $j$ running over $1$ to $r_2$ and so forth.  
 Fixing the  label $j$ fixes the column. 
 For example $\mu^i_1 $ has the label $i$ running 
 from $1$ to $c_1$,  $\mu^i_2 $ has $i$ running 
 from $1$ to $c_2$, and so forth. 
 
 The symmetric group $ S_{r_1} \times S_{r_2} \cdots \times S_{r_k}$
  acts on the  lower labels $j$ in  $ \mu^i_j$ keeping the 
    label $i$   fixed. We will call this $S_R$, the symmetric 
 group which acts along the rows.  
The symmetric group $ S_{c_1} \times S_{c_2} \cdots \times S_{c_{r_1} }$
 acts on the upper  labels $i$ in  $ \mu^i_j$ while keeping 
 the $j$ fixed. We will call this $S_C$, the symmetric group 
 which acts along the  columns.
 The group $S_C \times S_R $ is a subgroup of
 $S_B$. 
  The Young symmetrizer is a sum 
\eqn\youngsymm{  { 1 \over {| S_R |}  } { 1 \over { |S_C  | } }  
\sum_{\sigma \in S_R } \sum_{ \tau \in S_C } 
 (-1)^{\tau } \sigma \tau } 
$ (-1)^{\tau }$ is $+1$ if the permutation $\tau $ 
 is even and $-1$ if it is odd. 
 The factors in the denominator $ | S_R | $ and $ | S_C | $ 
 are the dimensions of $S_R $  and $S_C$ respectively, 
 i.e $ |S_R| =  r_1! \cdots r_{c_1} ! $ and 
  $| S_C |  =  c_1! \cdots c_{r_1} !$, respectively. 
 Equation \youngsymm\ expresses the fact that 
 we symmetrize over  the rows and antisymmetrize
 over columns.

 Now the space of traceless tensors still 
 has an action of $S_B$, and hence of $S_R \times S_C$. 
 The symmetrization operation applied to traceless 
 tensors $A(\mu^i_j)$ yields an irreducible 
 representation of $SO(2k + 1) $.
 The dimension of the representation can 
 be written neatly in terms of the following 
 quantities, 
\eqn\defs{\eqalign{&  l_i = r_i + k - i + { 1 \over 2 }  \cr 
                   &  m_i = k - i + { 1 \over 2 } \cr 
}} 
where $i$ runs from $1$ to $k$. 
  The dimension of the representation  is 
\eqn\dims{ 
 D( \vec r ) = \prod_{i< j } 
{ ( l_i^2 - l_j^2 ) \over ( m_i^2 -m_j^2 ) } \prod_{i} { l_i \over m_i}     
} 
and is derived in \fh, for example. 

For the representations constructed from 
tensor products of the vector of $SO(2k+1)$, the $ r_i$ 
 are all integers.  Vectors $\vec r$ with 
 half-integer entries 
 are weight vectors corresponding 
 to spinor representations.  
The dimensions of spinor representations can be obtained 
by substituting in \dims\ vectors $ \vec r $ with  half-integral values of 
$r_i$. For example the fundamental spinor 
is labelled by $ { \vec r }=  ( {1\over 2 }, { 1\over 2} , \cdots { 1\over 2 }  ) $. 
The $n$'th symmetric tensor power of the spinor 
is labelled by $ { \vec r } = 
   ( { n\over 2 } ,{ n\over 2 } ,\cdots, {n\over 2}  ) $.

\newsec{ The fuzzy 4-sphere } 

We here recall some facts about the fuzzy 4-sphere, 
 which is discussed in detail in \gkp\clt.  
For the fuzzy $S^4$,  the matrices $G^{\mu}$ satisfying 
$\sum_{\mu} G^{\mu}G^{\mu} = R^2$ act on vectors in  the irreducible
representation of $Spin(5)$ obtained by symmetrizing 
the $n$'th tensor power of the 4-dimensional spinor representation
\eqn\fuzfr{ G^{\mu} = \bigl( 
\Gamma^{\mu} \otimes 1 \otimes 1 \cdots \otimes 1 
                       + 1 \otimes \Gamma^{ \mu } \otimes 1 \cdots
\otimes 1 \cdots + 1 \otimes \cdots \otimes 1 \otimes \Gamma^{\mu }
\bigr)_{sym} } 
The index $\mu $ runs from $1$ to $5$. 
It is convenient to rewrite this 
as 
\eqn\fuzfr{ G^{\mu} = P_n \sum_{k} \rho_{k} ( \Gamma^{\mu } ) P_{n} } 
 where the right hand side is a set of operators 
 acting on the n-fold tensor product $V \otimes V \otimes  \cdots
 \otimes  V$. 
 The expression  $  \rho_{k} ( \Gamma^{\mu } ) $ is the 
 action of $ \Gamma^{\mu } $ on the $k$'th factor of the tensor 
 product. 
\eqn\defr{
 \rho_{k} ( A )  | e_{i_1} e_{i_2} \cdots e_{i_k} \cdots e_{i_n} > 
   =  A^{j_k}_{i_k}  | e_{i_1} e_{i_2} \cdots e_{j_k} \cdots e_{i_n} >
} 
The symmetrization operator  $P_n$ is given by 
 $ P_n = \sum_{\sigma \in S_n} { 1 \over n!} \sigma $
where $\sigma $ acts as : 
\eqn\defp{
 \sigma   | e_{i_1} e_{i_2} \cdots e_{i_n} > 
   =   | e_{i_ {\sigma(1)} } e_{i_{\sigma(2)}}  \cdots 
e_{i_{\sigma(n)}} >  }

 The symmetric $n$'th tensor power of the spinor, 
$ Sym( V^{\otimes n } ) $  is an irreducible 
 representation with dimension 
 $N =  { 1\over 6 } (n+1)(n+2)(n+3) $. 
 This can be checked by using  the weight vector 
 $ \vec r = { 1 \over 2 } ( n , n )$ in the dimension formula
 \dims.

 By taking various products of the $ G^{\mu}$ 
 we generate a class of $ N \times N $ matrices.
 Among these matrices are generators of $SO(5)$. 
 These take the form 
 \eqn\genso{ G^{\mu \nu } = 
\sum_{r=1}^{n}  \rho_r( [ \Gamma^{\mu}, \Gamma^{\nu}] ). }
 They act on the full set of matrices generated 
 by the $G^{\mu}$ through the commutator action. 
 As observed in \myers\gkp\ho\  the operators $G^{\mu} $
 and $G^{\mu \nu }$ close into the Lie algebra of $SO(6)$.  
 We will find the decomposition of the Matrix algebra
  into representations of $SO(5)$. 
 By summing up the dimensions of the representations 
 appearing in the decompostion we will find that 
 the $G^{\mu}$ matrices actually generate the full 
 set of $ N \times N$ matrices.

 This is easily accomplished by observing 
 that there is a simple relation between 
 the matrices generated by multiplying the 
 $G$'s and the Young diagram characterization of 
 the irreducible representations.  
 We will give a few examples to build up to 
 a correspondence between Young diagrams  and
 the matrices generated by the $G$'s.

 The $G^{\mu}$ transform as a vector of $SO(5)$, 
 the representation $ \vec r = ( 1,0) $.  
 Now consider products of two $G$'s :  
\eqn\twog{ G^{\mu_1} G^{\mu_2} = \sum_{s_1,s_2=1}^{n} 
\rho_{s_1}  ( \Gamma^{\mu_1} ) \rho_{s_2} ( \Gamma^{\mu_2} ) }  
 We can separate this sum into two  types 
 of terms, depending on whether $s_1$ is equal to $s_2$ 
 or not. The antisymmetrised product $  G^{\mu_1} G^{\mu_2} - 
G^{\mu_2} G^{\mu_1} $ only 
 contains terms where $ s_1=s_2$, i.e it is made 
 of terms like
\eqn\gmigmii{ \Gamma^{\mu_1}\Gamma^{\mu_2} \otimes 1 \otimes  \cdots
 \otimes 1 } 
with $ \mu_1 \ne \mu_2$. 
 A convenient way of writing the operator is
\eqn\convw{
 \sum_{s=1}^{n} \rho_{s} ( \Gamma^{\mu_1} \Gamma^{\mu_2} ). }  
 The set of linearly independent such matrices, 
 spanned by operators of the form given in 
 \convw\ with $ \mu_1 > \mu_2 $ is in $1-1$ correspondence 
 with vectors in the irreducible representation 
 labelled by the Young diagram  with
 one column of length $2$, i.e $\vec c = ( 2,0\cdots ) $ or 
 equvalently with row lengths given by the vector 
 $\vec r = ( 1,1 )$.  
The symmetric combination $ G^{\mu_1} G^{\mu_2} +  G^{\mu_2}
 G^{\mu_1}$ can be separated, in an $SO(5)$ invariant manner 
 into a traceless part and a 
 trace part. The trace part is   $ \delta_{\mu_1 \mu_2 } 
 ( G^{\mu_1} G^{\mu_2} +  G^{\mu_2}
 G^{\mu_1 } ) $. This is manifestly $SO(5) $ invariant, 
 so it is proportional to the identity matrix in the 
 $N$ dimensional irreducible representation.  

 The traceless symmetric part $  ( G^{\mu_1} G^{\mu_2} + G^{\mu_2}
 G^{\mu_1}  ) - \delta_{\mu_1 \mu_2}  G^{\mu_1} G^{\mu_2} $ 
 contains, from the sum \twog\ those expressions 
 with $ r_1 \ne r_2 $, and can be put in $1-1$ correspondence 
 with the irreducible representation of $SO(5)$ 
 which is associated with  the Young diagram with  
 $ \vec r  = ( 2,0)$. 

\subsec{ An $SO(5)$ covariant basis for $ N \times N $ matrix algebra  } 

 To develop further the connection between 
 the algebra generated by the $G^{\mu } $ and 
 the algebra of $ N \times N $ Matrices, we will first describe a 
 convenient basis for these matrices in terms of
 operators which correspond to irreducible representations 
 of $SO(5)$. 

 To each Young Diagram corresponding to an irreducible representation 
 of $SO(5)$ we will associate an operator built from $ \Ga $ matrices 
 acting in $ Sym( V ^{ \otimes n } ) $. As an example consider the
 diagram in the following figure. 

\medskip
\fig\ytabi{ 
{Figure 1 }  } 
{\epsfxsize1.4in\epsfbox{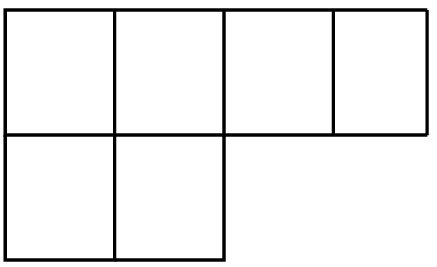} }

The above Young diagram corresponds to an operator of the form  
\eqn\dop{ \sum_{ { \vec s }}  \rho_{s_1}( \Ga \Ga ) \rho_{s_2}( \Ga
\Ga ) \rho_{s_3} ( \Ga ) \rho_{s_4} ( \Ga ) } 
The summation over ${\vec s }$ denotes a sum where the 
 indices $ s_1 \cdots s_4$ run from $1$ to $n$ 
while respecting the condition $ s_1 \ne s_2 \ne s_3 \ne s_4 $
(  Clearly such an operator only exists for $n \ge 4 $ ).
 For simplicity we have not included $SO(5)$ indices on the 
 $\Ga $ in \dop. We will now be more explicit. 
 The list of operators of the above form includes 
 expressions 
\eqn\diagop{
 \sum_{\mu^i_j} A[ \mu^1_1, \mu^2_1;    \mu^1_2, \mu^2_2 ; \mu^1_3; 
\mu^1_4 ] \sum_{\vec s  } \rho_{s_1} (\Ga^{\mu^1_1}
\Ga^{\mu^2_1}) \rho_{s_2} ( \Ga^{\mu^1_2} \Ga^{\mu^2_2} ) 
\rho_{s_3} ( \Ga^{\mu^1_3} ) \rho_{s_4} ( \Ga^{\mu^1_4} ) } 
The tensor $ A $ is traceless and has the Young Diagram symmetry, i.e 
it is symmetric under permutations of the index $j$ 
of $\mu^i_j $ for fixed $i$, and is antisymmetric under 
permutations of the index $i$ for fixed $j$.   
The number of operators of this form is the dimension of the
representation of $SO(5)$ associated with the Young diagram 
with row lengths given by $ \vec r = (4,2) $,  which is $220$ using
\dims.  

In general for a Young diagram with row lengths 
$ \vec r = (r_1, r_2 ) $, we have 
\eqn\frmg{\eqalign{ 
 \sum_{\mu^i_j} A[ \mu^i_j]  \sum_{ \vec s } &\rho_{s_1} (\Ga^{\mu^1_1}
\Ga^{\mu^2_1}) \rho_{s_2} (\Ga^{\mu^1_2} \Ga^{\mu^2_2})   \cdots 
 \rho_{s_{r_2}}  ( \Ga^{\mu^1_{r_2} } \Ga^{\mu^2_{r_2}} ) \cr 
& \rho_{s_{r_2 +1 }} ( \Ga^{\mu^1_{r_2+1}} ) \cdots 
\rho_{s_{r_1 }} ( \Ga^{\mu^1_{r_1 } } )  \cr }} 
The number of $ \mu$ indices is equal to the number 
of boxes $B$ in the Young diagram, where  $  B = r_1 +r_2 $. 
The tensor $A$ is traceless and has the symmetry of the Young diagram, 
 i.e it is symmetric in the index $j$ and antisymmetric in the index 
 $i$.

 Representations of $SO(5)$ have Young diagrams 
 with at most two rows. The number of rows, in the 
 correspondence with operators we described above, 
 maps to the largest number of $ \Gamma $ matrices acting on a single $V$. 
 In listing the independent operators in $Mat_N( C ) $, 
 we do not need more than two $\Ga$ acting on 
 any factor $V$ since the identity $ \Ga^5 = \Ga^1 \Ga^2 \Ga^3 \Ga^4 $
 can be used to write any product of more than two $\Ga $ 
 matrices in terms of a product with fewer $\Ga$'s. 
 Hence  the class of operators we need to consider
 in a basis for $Mat_N$ matches conveniently with the 
 representations of $SO(5)$.  Similarly we can understand the need
 for the tensor $A$ to be traceless in giving a basis 
 of independent operators in the Matrix algebra. 
 If we have an operator of the form 
 $ \sum_{\mu} \sum_{\vec s } 
\rho_{s_1}( \Ga^{\mu}  \Ga^{\nu}  ) \rho_{s_2} ( \Ga^{\mu} ) $, 
 for example, we can use the identity 
 $ \sum_{\mu} \Ga^{\mu} \otimes \Ga^{\mu} = 1 \otimes 1 $
 in $ Sym( V \otimes V) $, to show that it is not independent 
 of operators involving only one $\Ga$.  
 The antisymmetry in the $j$ index labelling different 
 $\Ga$ matrices acting in the same $V$ is  
 understood since any $\Ga$ matrix squares to the identity. 
 The symmetry along the rows 
  is actually automatic for the operators \frmg\  since the different $s$ 
 indices can be renamed without changing the operator.

 The complete set of irreducible representations 
 of $SO(5)$ in $ Mat_{N}(C)$   is given by 
\eqn\setrep{  n \ge r_1 \ge r_2 \ge 0 } 
  We sum up the dimensions as given by  
 \dims. First summing over the length of the second row, 
 to get a function of the first row length, we get
\eqn\stpsm{  D(r_1) \equiv \sum_{r_2=0}^{r_1} D(r_1,r_2) = { 1 \over 12}
 ( r_1 +1)^2 (r_1+2)^2 (2r_1+3)^2 }    
Doing now the sum over the first row length : 
\eqn\stpisum{ \sum_{r_1=0}^{n} D(r_1) = { 1 \over 36 }
 ( n+1)^2(n+2)^2(n+3)^2 }
This is exactly equal to the square of $N$. 
 This proves that the above operators \frmg\ give a complete 
$SO(5)$ covariant basis 
 for the Matrix algebra, $ Mat_N( C ) $. 
We may summarize this result as : 
\eqn\sfsum{   Mat_N( C ) = \oplus_{n \ge r_1 \ge r_2 \ge 0 }
 V_{r_1,r_2 }  .} 
Here $V_{r_1,r_2 } $ refers to operators which transform 
in the irreducible representation of $SO(5)$ labelled by the
 row lengths $\vec r = (r_1,r_2 )$. The explicit form 
 of such operators corresponding to a given representation 
 is given by \frmg.

\subsec{ Basis and products of $G^{\mu}$ }  
 Consider a set of $B$ copies of the matrices  $G^{\mu}$. 
 We can form  sums of products of these matrices 
 using the traceless tensors with symmetry specified 
 by Young diagrams $Y$ with $B$ boxes and $\vec r = (r_1,r_2 )$. 
 They take the form 
\eqn\symts{  \sum_{\mu^i_j} A ( \mu^i_j ) \prod_{i,j}  G^{\mu^i_j}. }  
 This set of operators transforms according to the Young diagram
 $Y$. We let $r_1,r_2$ range in the interval given by \setrep. 
 Since we showed above that the dimensions of precisely 
 this set of representations adds up to $N^2$,  
 we know that these products of $G^{\mu} $ also form a basis of the
 algebra. In fact we know that the products of 
 $G^{\mu} $ corresponding to a Young diagram has to be proportional 
 to the operator of the form \frmg\ associated with the 
 same Young diagram. The constant of proprtionality 
 can be determined by acting on one state in $Sym( V^{\otimes n} ) $. 
 This can be used to simplify calculations involved in the study of 
 polarization of branes. 
 
\subsec{ Projection from Matrix Algebra to Fuzzy Spherical harmonics }  

 The fuzzy sphere algebra we have discussed so far 
 $ \chAn( S^4 )$, which is isomorphic to 
 $ Mat_N(C )   $, contains,   in the large $n$
 limit, all  the $SO(5)$ representations labelled by the 
 row lengths $r_1$ and $r_2$, with unit multiplicity. 
 Only the representations 
 with $r_2=0$ give spherical harmonics 
 of the classical sphere. So just looking at the
 algebra $\chAn(S^4) $ is not the adequate non-commutative 
 structure which leads to the classical sphere in the large $n$ 
 limit. Rather we need to consider the an 
 algebra $\cAn ( S^4) $ together with the equation $ r_2 =0$. 
 Any $N \times N $ matrix  $A$ can be written as a linear combination 
 of operators transforming in symmetric representations with
 $r_2 =0 $, which we call 
 $A_+$, and a linear combination involving non-symmetric
 representations with  $r_2 \ne 0$ which we call 
 $A_-$ : 
\eqn\dcmp{ A = A_+ + A_-. } 
 We define a projection $P$ which annihilates the $A_-$ : 
\eqn\prja{ P(A) = A_+ . } 
 If we have two elements $A$ and $B$, which satisfy 
 $ P(A) = A $ and $ P(B) = B$, then in general the matrix  product  
  $ A B$ does not have to be symmetric. We can express this as : 
\eqn\prodpm{ AB = (AB)_+ + (AB)_- } 
$(AB)_+$ denotes matrices which transform 
 as representations with $r_2=0$. $(AB)_-$ denote 
 matrices which tramsform in representations with $r_2 \ne 0$. 
However we can define 
 a product which closes on the symmetric representations 
 by just projecting the matrix product  
\eqn\newpd{ A \bullet B  = P ( A B ) = (AB)_+ } 
We show, in a  later section,  that this is not an associative 
 product, and that the non-associativity disappears 
 in the large $n$ limit so that we indeed recover the 
 classical sphere. Here we just observe a simple numerical 
 reason for suspecting non-associativity. 
  While a sum of representations with non-zero 
 $r_1$ and $r_2$ adds up as in \stpisum\ to an absolute square 
 and hence can be identified with a Matrix algebra, a sum 
 with non-zero $r_1$ only, up to $n$, 
 gives $ { 1\over 12} (n+2)^2 (n+1) (n+3)$. 
 This is not an absolute square. 
 Now under rather general grounds, whenever we have 
 an irreducible representation of an associative algebra 
 over the complex numbers, 
 the  representatives of the algebra give a 
 complete basis for the matrix algebra. This is Burnside's 
 theorem, described for example in \gdwa. 
  The symmetric representations, added with 
 unit multiplicity,  do not add up to the 
 dimension of any matrix algebra so we do not 
 expect them to form  a finite Matrix algebra over $C$ , without 
 knowing any details of explicit fuzzy sphere constructions 
 of the kind we described. 

 The projection defined in \prja\ and the product in \newpd\ 
 can be re-expressed 
 as follows. The quadratic and cubic casimirs 
 of $SO(5) $  can be written in terms of 
 products of the generators of  $SO(5)$ $G^{\mu \nu }$.
 In any irreducible representation, 
 these are numbers which are functions of $r_1$ and $r_2$. 
 The explicit formulae are given for example 
 in \baracz.  
 We can invert this relation in order to write 
 a formula for $r_2$ in terms of the Casimirs, and hence 
 in terms of a series in  $G_{\mu \nu}$. These  can be considered 
 as differential operators on the algebra $\chAn(S^4) $, and they act by
 commutators. 
 Hence $r_2$ can be considered as a differential operator 
 on $\chAn(S^4) $, which we write as $ {\hat r }_2$. 
 $A_+$ in \prja\ is the kernel of this differential operator. 
 A useful way to write the projected
 product, 
 is to multiply $AB$, the ordinary product, with spherical 
 harmonics $ Y_{r_1,r_2=0}^*$,  take a trace to pick out the
 coefficient of $Y(r_1,r_2=0)$ in the expansion of $AB$,   
 and sum over $r_1$ 
\eqn\newpdi{ P(AB) = \sum_{r_1} TR( AB Y_{(r_1,0)}^* ) Y_{(r_1,0)}.  }

\newsec{Higher even spheres }

The construction we described above for the fuzzy
 four-sphere admits a simple generalization to higher even spheres. 

\subsec{ The fuzzy 6-sphere } 

 The arguments we described in detail for the fuzzy $S^4$ 
 generalize straightforwardly to the fuzzy $S^6$. 
  Now we consider operators 
 $G^{\mu}$ acting on the symmetrized tensor space 
 $ Sym ( V^{\otimes n } ) $ where $V$ is the 
 fundamental spinor with weight $ \vec r  = ( { 1 \over 2 }, { 1 \over
 2 } , { 1 \over 2 } ) $. 
 The symmetrized tensor product of the spinor has weight 
 $ \vec r  = ( { n \over 2 }, { n \over 2 },  { n \over 2 } ) $. 
The action of $G^{\mu}$ is again given 
 by 
\eqn\gmac{ G^{\mu} = P_n \sum_{r=1}^{n} \rho_r( \Gamma^{\mu} ) P_n }   
 Using the dimension formula \dims, we get 
\eqn\dmrpsx{ N \equiv Dim ( \cR )  = 
{ 1 \over 360 } ( n + 1 ) ( n +2 ) ( n+3)^2 ( n+4 ) ( n+5 ).  } 

 Now the tensor representations are given by Young diagrams 
 with three rows, labelled by $ \vec r = ( r_1, r_2,r_3 ) $, 
 with $ r_1 \ge r_2 \ge r_3 $. For fixed $n$, the operators 
 that appear in Matrices over $\cR $ also obey the restriction 
 $ r_1 \le n $.   
 The map between representations and matrices acting on $ \cR $
 analogous to \frmg\ is given by 
\eqn\frmgi{\eqalign{ 
 \sum_{\mu^i_j} A[ \mu^i_j]  \sum_{ \vec s } &\rho_{s_1} (\Ga^{\mu^1_1}
\Ga^{\mu^2_1} \Ga^{\mu^3_1}  )  \cdots 
 \rho_{s_{r_3}}  ( \Ga^{\mu^1_{r_3} } \Ga^{\mu^2_{r_3}}
 \Ga^{\mu^3_{r_3}} ) \cr 
& \rho_{s_{r_3 +1 }} ( \Ga^{\mu^1_{r_3+1}}  \Ga^{\mu^2_{r_3+1}}  ) \cdots 
 \rho_{r_2 } ( \Ga^{\mu^1_{r_2}}  \Ga^{ \mu^2_{r_2} }  ) \cdots \cr 
& \rho_{s_{r_2+1 }} ( \Ga^{\mu^1_{r_2+1 } } ) \cdots 
     \rho_{s_{r_1 }} ( \Ga^{\mu^1_{r_1 } } )  \cr }} 
The equation  describing products of $G_i$ 
 which correspond to a given representation 
take the same form as \symts, with the index 
 $i$ running from $1$ up to  a maximum of $3$ and the index $j$ 
 running from $1$ up to a maximum of $n$, and $A$ being a
 traceless tensor with the Young symmetry.  

 The number of operators corresponding to a 
 given Young diagram is the dimension of that representation. 
 The dimension $D(r_1,r_2,r_3 )$ 
 is given by \dims. 
 Doing the sums using Maple, we find, after the sum over $r_3$
\eqn\smi{\eqalign{ 
 D(r_1,r_2) & \equiv 
 \sum_{r_3=0}^{r_1} D(r_1,r_2,r_3)\cr 
 & = { 1 \over 4320 } ( 2r_2 +3 ) 
 (r_2 +2)^2 ( r_2 +1 )^2 (2r_1+5) \cr 
&  ( r_1+r_2+4) (r_1-r_2+1) ( 3r_1^2 +
 15 r_1 - r_2^2 - 3 r_2 + 18 ) \cr } }  
 After the sum over $r_2$ we have 
\eqn\smii{\eqalign{  D(r_1) & \equiv  \sum_{r_2=0}^{r_1} D(r_1,r_2)
 \cr 
                   & = { 1 \over 43200} (2r_1 +5) ( 2r_1^2 + 10r_1 + 15 ) 
                      ( r_1 +4)^2 ( r_1 + 3 )^2 ( r_1 + 2)^2
 ( r_1+1)^2 \cr }}
 Finally we do the sum over 
$r_1$ and find 
\eqn\smiii{ D = \sum_{r_1} D(r_1) = N^2 }  
where $N$ is given in \dmrpsx. 
   
If we just sum over the representations 
with a non-zero first row, with the length of teh first row 
 extending from $1$ to $n$, we get 
$ { 1\over 360} ( n+3)^2 (n+1)(n+2)(n+4)(n+5) $, which is not 
 an absolute square.

\subsec{ Fuzzy $S^8$ } 

 In this case we take the fundamental spinor
 of $SO(9)$ and consider  the $n$'th symmetric tensor 
 power. The dimension of this representation 
 is $N = { 1 \over 302400 } (n+1)^2
  (n+2)^2 (n+3)^4 (n+4)^4 (n+5 )^4 ( n+6)^2 (n+7)^2 $. 

The $G^{\mu} $ are $\Gamma$ matrices 
 acting in these representations. 
 We can associate independent matrices 
 generated by multiplying $G^{\mu}$
 with traceless tensors associated with symmetry described 
 by Young diagrams 
 $Y$ having row lengths $ ( r_1, r_2, r_3, r_4
 )$, with $ n \ge r_1 \ge r_2 \ge r_3 \ge r_4
 \ge 0 $. By summing ( using Maple )  over all such representations 
 with $r $'s in this range we again get 
 precisely $N^2$.

\newsec{Review :  Even orthogonal groups $SO(2k)$ } 

 We need to review some properties of the 
 $SO(2k)$, which can be found in \fh\ for example. 
 The  tensor representations are labelled by 
 integer valued vectors $ \vec r  = ( r_1, r_2
 .. , r_k ) $. 
 The difference from $SO(2k+1 )$ is that
  $ r_k$ can be either positive or negative. 
  In the case of $SO(4) = SU(2) \times SU(2) $, 
  we have 
 \eqn\lfrt{\eqalign{&  2j_L = r_1 + r_2 \cr 
                   &  2j_R = r_1 - r_2 \cr }} 
 When $r_2 < 0 $, $2j_L < 2j_R $.  
 For example,  the representation with 
  $ \vec r = (1,1) $ corresponds to 
  self-dual anti-symmetric tensors, and the representation 
 with $ \vec r  = (1,-1) $ corresponds to anti-self-dual tensors.  
  In general we will refer to representations associated to 
 vectors where $r_k > 0 $ as self-dual, and those 
 with $r_k < 0 $ as anti-self-dual. Two representations 
 which are related by a change of sign of $r_k$ are 
 conjugate.  
 
 To write a formula for the dimensions 
 of representations we define 
\eqn\defse{\eqalign{&  l_i = r_i  + k - i  \cr
                   &  m_i =  k - i \cr }}
The dimension is 
\eqn\dimse{ D(l_i ) = \prod_{i \le j } { ( l_i^2 - l_j^2 )\over (m_i^2
 - m_j^2 ) }  } 
The construction of irreducible representations 
 with $r_k \ne 0 $ proceeds by applying the Young symmetrizer 
 to traceless tensors. For $r_k > 0 $, we apply to each
 antisymmetrized 
 product of $k$ tensors ( corresponding to the columns of length 
 $k$ in the Young Diagram ), a $P_+$ projector, where 
$ P_+ = { 1 \over 2 } ( 1 + \Gamma^{ 2k + 1 } ) $.  
 For $r_k < 0 $, we apply the $P_-$ projector in the same way, 
 where $ P_+ = { 1 \over 2 } ( 1 - \Gamma^{ 2k + 1 } ) $.

\newsec{ The fuzzy 3-sphere } 

 For the fuzzy 3-sphere we are working with 
 matrices acting in a reducible representation 
 $ \cRp \oplus \cRm $. The weights of $\cRp$ are 
 obtained by adding ${ (n+1) \over 2} $ copies of
 $ \vec r = ( \haf, \haf )$ and $ {(n-1) \over 2}  $ copies 
 of $ \vec r =  ( \haf, - \haf ) $, giving a total 
 weight of $ \vec r = ( { n \over 2 } , \haf ) $. 
 The representation $\cRm$ has weights 
 $ \vec r =   ( { n \over 2 } , - \haf )$.  
 The dimensions are $ D(R_+) = D(R_-) = { (n+1 ) (n+3) \over 4 }$, 
 as can be checked using the formula \dimse. 
 Alternatively we can exploit the isomorphism 
 $ SO(4) = SU(2) \times SU(2) $ which gives the relation \lfrt.

 The coordinates of the fuzzy three-sphere were defined 
 in \zr\ by the equation
\eqn\coord{  {\hat G}_i = \Pr  G_i \Pr }
where $ \Pr = \Prp + \Prm $ and acts on  $ Sym( V^{\otimes n } ) $. 
 $V$ is a reducible representation $ V = V_+ \oplus V_- $, 
 where $V_+ $ has weights $ \vec r = ( \haf, \haf )$ or  
$( 2j_L , 2j_R) = ( \haf, 0 ) $ 
 and $V_- $ has weights $ \vec r = ( \haf, - \haf )$ or 
$( 2j_L , 2j_R) = ( 0 , \haf ) $. 
 $V$ is an irreducible representation of $Spin(5)$ which is used 
 for the even fuzzy sphere $ \chAn( S^{4} )$.   
 $\Prp$ is an operator 
  which projects to $ \cRp$, the 
 irreducible representation of $ Spin(4) $ labelled 
 by $( 2j_L , 2j_R) = ( { (n+1)  \over 2 }, { (n-1 ) \over 2 } ) $. 
 The operator    $\Prm$ projects to  
 $( 2j_L , 2j_R) = ( { (n+1)  \over 2 }, { (n-1 ) \over 2 } ) $.
 The explicit expression is 
$ \Prp = ( ( P_+)^{ \otimes  {n+1 \over 2 }}  P_-^ { \otimes {n-1
 \over 2 }}  )_{sym} $, where the subscript denotes symmetrization. 
 For example, in the case $n=3$, 
 $$ \Prp = P_+ \otimes P_+ \otimes P_- ~~+ ~~   P_+ \otimes P_-
\otimes P_+ ~~+ ~~  P_-  \otimes P_+ \otimes P_+. $$
 
\subsec{ $SO(4)$ covariant basis for Matrices acting on $\cR$  } 
 To  obtain the representations of 
 $SO(4)$ which are contained in Matrices 
 acting on $\cRp  \oplus \cRm $, we separate the problem 
 into two parts. We first consider 
 matrices  taking $\cRp $ to $\cRp$,  which we denote 
 as $ End( \cRp )$. Next we consider matrices 
 mapping $\cRp$ to $\cRm$ which we denote 
 as $Hom( \cRp , \cRm )$.  The decomposition 
 of $ End( \cRm )$ follows from that of 
 $ End( \cRp ) $ by  changing $P_+$ to $P_-$. 
 Similarly the decomposition of 
 $Hom(\cRm, \cRp )$ follows from that of 
 $Hom(\cRp, \cRm) $. 

 Let us consider then matrices in $End( \cRp )$. 
 They correspond to Young Diagrams with even numbers 
 of boxes $B$. 
 We first consider operators transforming 
 in self-dual representations. These are labelled 
 by $\vec r = (p_1 + p_2, p_1 )$ , with $p_1 > 0 $ and $p_2 \ge 0$.  
  The schematic  form of the operators corresponding to the above 
 ``self-dual Young Diagram''
\eqn\slfdl{ 
\sum_{ \vec s, \vec t}  \rho_{s_1} ( \Gamma \Gamma P_+ ) \rho_{s_2}
 ( \Gamma \Gamma P_+ ) \cdots \rho_{s_{p_1}} ( \Gamma \Gamma P_+) ~~
 \rho_{t_1} ( \Gamma ) \rho_{t_2} ( \Gamma ) \cdots \rho_{t_{p_2} }
 ( \Gamma  )  } 
 The sum over $(\vec s, \vec  t)$ 
denotes a sum where $ s_1 \ne s_2 \cdots \ne s_{p_1}
 \ne t_1 \cdots \ne t_{p_2 } $ and all the $s$ and $t$ range over 
 $1$ to $n$. We have suppressed the $SO(4)$ indices on the 
 $ \Gamma $, to avoid writing a more cumbersome expression. 
 It is understood that the $ \Ga$ carry $SO(4) $ indices 
 which are contracted with a traceless tensor with Young symmetry 
 as in \frmg.  
 All the operators corresponding to the 
 $p_1$ columns of length $2$ have $P_+$ projectors 
 attached to them. This guarantees that the operator transforms 
 according to a self-dual representation. 
 Recall that $\cRp$ is obtained from $Sym(  V^{\otimes n } )  $ by acting 
 with strings of ${ (n +1) \over 2 }$ copies of $P_+$, 
 and ${(n-1)\over 2 } $ copies of $P_-$. 
 Of the $p_2$ copies $ \rho( \Gamma )$ acting on $ \cRp $, 
 let $p_2^+ $ denote the number that act on a $P_+$ 
 and $p_2^-$ the number that act on a $P_-$. And let $ p_3^+ $
 denote the remaining number of $P_+$. 
 By adding up the number of $P_+ $ that the operator \slfdl\ is acting
 on, we get 
\eqn\incom{ p_1 + p_2^+  + p_3^+ = { (n+1) \over 2  } } 
 and adding the number of $P_+$ operators that result from this action 
\eqn\outpls{ p_1 + p_2^- + p_3^+ =  { (n+1) \over 2  } } 
These two equations imply that 
\eqn\reslt{ p_2^+ = p_2^-.  } 
Let us define $ k = p_2^+ = p_2^- $. 
  Without loss of generality, then, we can 
 write  for the self-dual operators 
\eqn\slfdli{\eqalign{&  \sum_{{\vec s} , {\vec t} }\rho_{s_1} ( \Gamma \Gamma P_+ ) \rho_{s_2}
 ( \Gamma \Gamma P_+ ) \cdots \rho_{s_{p_1}} ( \Gamma \Gamma P_+) 
 \rho_{t_1} ( \Gamma P_+) \rho_{t_2} ( \Gamma P_+ ) \cdots \rho_{t_{k}  }
 ( \Gamma P_+ ) \cr 
& \rho_{t_{k+1} } ( \Gamma P_-) \rho_{t_{k+2} } 
( \Gamma P_- ) \cdots \rho_{t_{2k}}
 ( \Gamma P_- ) \cr }}
  The allowed values of $k$ range from 
 $ 0 $ to $ {(n-1) \over 2 } $, since there are no more than 
 $ {(n-1) \over 2 } $ copies of $P_-$. 
For fixed $k$, 
  the allowed  values of $p_1$ range then from 
  $0$ to $ { (n+1) \over 2 } - k  $. Doing the sum over the
appropriate
 representations we get : 
\eqn\slfdld{ \sum_{k=0}^{(n-1)\over 2 } \sum_{p_1=0}^{ { (n+1)\over 2 } -
k} ( 2 p_1 + 2 k + 1 ) ( 2k + 1)  =
 {1 \over 96} ( n+1) ( 3n^3 + 41 n^2 + 97 n + 51 ) } 
 We have included, in the above, a term with $p_1 =0 $, which  
 corresponds to representations which are not self-dual 
 or anti-self-dual. 

 For anti-selfdual representations we have 
 $ \vec r = ( p_1 + p_2 , -p_1 )$. 
Now the operators are of the form 
\eqn\aslfdl{ 
\sum_{{ \vec s}, {\vec t} }  \rho_{s_1} ( \Gamma \Gamma P_- ) \rho_{s_2}
 ( \Gamma \Gamma P_- ) \cdots \rho_{s_{p_1}} ( \Gamma \Gamma P_-) 
 \rho_{t_1} ( \Gamma ) \rho_{t_2} ( \Gamma ) \cdots \rho_{t_{p_2} }
 ( \Gamma  ). } 
Again we let $p_2^+$ and $p_2^-$ be the number 
of single $ \Gamma $ which act on $P_+$ and $P_-$ respectively, 
 and $p_3^- $ the number of remaining $P_-$. 
 Adding up the the number of $P_-$ that are acted on by the above 
 operator we get : 
\eqn\pmin{ p_1 + p_2^- + p_3^- = { (n-1) \over 2 } }
 and adding the number of $P_-$ that come out 
 we have: 
\eqn\pminout{ p_1 + p_2^+ + p_3^- = { (n-1) \over 2 } }
 This gives $ p_2^+ = p_2^- \equiv k $.  
Now $p_1$ can range from 
 $1$ to ${ (n-1) \over 2 }$, and  $k$ ranges from $0$ to 
 $ { (n-1) \over 2 } - p_1$. Performing the sum we get : 
\eqn\aslfd{ \sum_{p_1=1}^{(n-1)\over 2 } \sum_{k=0}^{ { (n-1)\over 2 } -
p_1 } ( 2 p_1 + 2 k + 1 ) ( 2k + 1)  =
 { 1 \over 96 } ( n-1) (n+1) ( 3n^2 + 4n + 3 ).  } 
     
 Adding up the dimensions  in \aslfd\ and \slfdld\ the self-dual and the antiself-dual
 representations we get $ { 1\over 16 } (n+1)^2 (n+3)^2 $, 
 which is exactly the dimension of the space of matrices
 mapping $\cRp$ to $\cRp$. 
 The matrices mapping $ \cRm$ to $ \cRm$ can be obtained 
 exactly as above by exchanging the role of $P_+$ with that of 
 $P_-$.

\subsec{ Decomposing $Hom(\cRp , \cRm )$ into representations of $SO(4)$ } 

Now we consider matrices which map $ \cRp $ to $ \cRm$.
We give here a complete basis for these Matrices. 

Operators transforming according to self-dual representations 
take the form: 
\eqn\hmsf{\eqalign{ 
\sum_{s,t}  ~~~~~~~&\rho_{s_1} ( \Gamma \Gamma P_+ ) \rho_{s_2}
 ( \Gamma \Gamma P_+ ) \cdots \rho_{s_{p_1}} ( \Gamma \Gamma P_+) \cr 
& \rho_{t_1} ( \Gamma ) \rho_{t_2} ( \Gamma ) \cdots 
  \rho_{t_{p_2}} ( \Gamma  ) \cr } } 
Of the $p_2$ copies $ \Gamma$ let $p_2^+ $ act on $P_+$ 
and let $p_3^+$ be the number of remaining $P_+$. 
The total number of $P_+$ that the above operator 
is acting on is, therefore, 
\eqn\ttlp{ p_1 + p_2^+ + p_3^+ = { (n+1) \over 2 } }
Here we used the fact that the projector $\Prp$, which projects 
$ Sym( V^{\otimes n} )$ to $ \cRp$ contains $ { n+1 \over 2 } $ 
 copies of $P_+$.   
The total number of $P_+$ which result from the action 
is
\eqn\ttlm{ p_1 + p_2^- + p_3^+ = { (n-1) \over 2 }. }
Here we used the fact that $\Prm$ has ${ (n-1) \over 2}$ copies 
of $P_+$.  
These two equations imply $ p_2^+ = p_2^- + 1 $. 
Letting $p_2^- = k$, we have $p_2 = p_2^+ + p_2^- = 2k +1 $. 
We can now write the operator in \hmsf\ without loss of 
generality as 
\eqn\hmsfi{\eqalign{& 
\sum_{\vec s, \vec t}  \rho_{s_1} ( \Gamma \Gamma P_+ ) \rho_{s_2}
 ( \Gamma \Gamma P_+ ) \cdots \rho_{s_{p_1}} ( \Gamma \Gamma P_+) \cr 
& \qquad\qquad \rho_{t_1} ( \Gamma P_+ ) \rho_{t_2} ( \Gamma  P_+) \cdots 
  \rho_{t_{k+1}} ( \Gamma P_+    ) \cr 
& \qquad \qquad \rho_{t_{k+2} } ( \Gamma P_- ) 
 \rho_{t_{k+3} } ( \Gamma  P_-) \cdots 
  \rho_{t_{2k+1}} ( \Gamma P_-   )
\cr } } 

From \ttlm\ the largest value of $p_1$ is ${ n-1 \over 2 }$, 
since $p_2^- \ge 0 $ and $ p_3^+ \ge 0$. The self-dual representations  
have $p_1 \ge 1$. The representations with $p_1=0$ are neither self-dual nor 
anti-self-dual. We will count them here by incuding the term with 
 $p_1=0$ along
with the self-dual reps.
It also follows that, for fixed $p_1$, $k = p_2^-$ is allowed to range from 
$0$ to $ { n-1 \over 2 } - p_1$. The number of Matrices which are of
this form can be obtained by adding up the dimensions 
of the corresponding representations. Here $r_1 = p_1 + 2k +1 $, 
and $r_2 = p_1 $, so that $2j_L = 2p_1 + 2k +1 $ and $2j_R = 2k +1 $.     
The contribution from these representations,   denoted by 
$D_+$ is therefore : 
\eqn\dpls{\eqalign{ 
  D_+ 
 &= \sum_{p_1 = 0 }^{(n-1)\over 2 } \sum_{k=0}^{ { (n-1)\over 2 } - p_1
  }  ( 2p_1 + 2k + 2 ) (2k +2 ) \cr
 &= { 1 \over 96 }  ( 3 n +5 ) ( n + 5 ) 
( n +  3 ) ( n + 1) \cr }}

Now consider the contribution from anti-self-dual 
representations corresponding to operators of the form: 
\eqn\hmssf{\eqalign{& 
\sum_{{\vec s} , {\vec t} }  \rho_{s_1} ( \Gamma \Gamma P_- ) \rho_{s_2}
 ( \Gamma \Gamma P_- ) \cdots \rho_{s_{p_1}} ( \Gamma \Gamma P_-) \cr 
& \qquad\qquad \rho_{t_1} ( \Gamma ) \rho_{t_2} ( \Gamma ) \cdots 
  \rho_{t_{p_2}} ( \Gamma  ) \cr } } 
Defining $p_2^+,p_2^- $ and $p_3^+$ 
as in the previous paragraph, 
and adding up the number of $P_+$ in the incoming 
 state we have  : 
\eqn\incom{ p_2^+ + p_3^+ = { n+1 \over 2 }, }  
and from the outgoing state  
\eqn\outgo{  p_2^- + p_3^+ = { n-1 \over 2 }  } 
This gives $p_2^+ = p_2^- + 1 $ as before, and we define $k \equiv
p_2^-$. Now the operator can be rewritten 
\eqn\hmsfi{\eqalign{& 
\sum_{s,t}  \rho_{s_1} ( \Gamma \Gamma P_- ) \rho_{s_2}
 ( \Gamma \Gamma P_- ) \cdots \rho_{s_{p_1}} ( \Gamma \Gamma P_-) \cr 
& \qquad\qquad \rho_{t_1} ( \Gamma P_+ ) \rho_{t_2} ( \Gamma  P_+) \cdots 
  \rho_{t_{k+1}} ( \Gamma P_+    )  \cr 
& \qquad \qquad \rho_{t_{k+2} } ( \Gamma P_- ) 
 \rho_{t_{k+3} } ( \Gamma  P_-) \cdots 
  \rho_{t_{2k+1}} ( \Gamma P_-   )
\cr } } 
Since we already counted representations with $p_1 =0 $ above, 
the allowed range of $p_1$ is from $1$ to ${n-1 \over 2 } $. 
The upper bound comes from the fact that  
 there are no more than ${ n-1 \over 2 } $ copies of $P_-$ in
$ \cRp$. For fixed $p_1$, $k$ can range from $0$ to $ { n-1\over 2 } -
p_1 $. 
The contribution from anti-self-dual representations, denoted by $D_-$
is  
\eqn\antsf{\eqalign{ 
  D_- &= \sum_{p_1=1}^{ {n+1 \over 2 } } \sum_{p_2=0}^{ {n-1 \over 2 }
- p_1} ( 2 p_1 + 2 p_2 + 2 ) ( 2p_1 + 2 ) \cr 
    &= {1 \over 96} (n-1)(3n+7)(n+3)(n+1) \cr }}

Adding up the expressions from \antsf\ and \dpls\ we find 
\eqn\totdtr{ D = D_+ + D_- = {1\over 16} (n+1)^2 (n+3)^2 }
This shows that the matrices described in 
\hmsfi\ and \hmsf\ give a complete $SO(4)$ covariant 
 basis for $Hom( \cRp, \cRm )$, the matrices acting from 
 $\cRp$ to $\cRm$. By exchanging $P_+$ with $P_-$ we can obtain
 $ Hom( \cRm, \cRp )$.

\subsec{ Relation between algebra generated by coordinates and Matrix
  Algebra } 

 In the case of $ \chAn( S^4 )$
  the algebra generated by the coordinate matrices  
 $ G^{\mu }$ is isomorphic to a Matrix algebra. 
  For the fuzzy three-sphere, this is not the case. 
 The coordinate matrices $\hat G^i$ do not generate 
  the matrix algebra $ Mat_N ( C ) $. They only generate a 
 sub-algebra. In the discussion of $Mat_N (C) $, 
 we considered $4$ subalgebras $ End( \cRp ) $, $ End( \cRm ) $, 
 $ Hom ( \cRp, \cRm ) $ and $ Hom ( \cRm, \cRp ) $. 
 Each has dimension $ { N^2 \over 4 }$. 
  Among the matrices in $ End ( \cRp )$, we have, for generic $n$ ( i.e
  $ n > 1  $ ),  
\eqn\twsamp{\eqalign{
&  \Prp \sum_{s} \rho_s ( G^{ij} P_+ ) \Prp \cr 
&  \Prm \sum_{s} \rho_s ( G^{ij} P_- ) \Prp \cr }}
Among $ End( \cRm ) $ we have  
 \eqn\twsampi{\eqalign{
&  \Prm \sum_{s} \rho_s ( G^{ij} P_+ ) \Prm \cr 
&  \Prm \sum_{s} \rho_s ( G^{ij} P_- ) \Prm \cr }}
We have used the notation $G^{ij} = [ \Ga^i, \Ga^j ] $. 
The matrices generated by $ {\hat G}^i $ only include  
 the first operator in \twsamp\ and the second in \twsampi. 
The operator $ \sum_{s} \rho_s ( \Ga^i ) $ acting on  
$\cRp $ can be written as a sum of $ \sum_{s} \rho_s ( \Ga^i P_+  ) +
\sum_{s} \rho_s ( \Ga^i P_-   )  $. The first operator  
 maps states from $\cRp $ to $ \cRm$. The second maps  
 states in $ \cRp$ to states in 
 the representation $ (2j_L, 2j_R ) = ( { n+3 \over 2 }, { n-3 \over 2
  } )$. These states are projected out by the $\Pr $ in the definition 
 of $ { \hat G}^i $. This is the reason why  operators in the  
  second line of \twsamp\ are in $ End( \cRp )$ but cannot  
  be generated by $ {\hat G}^{i}$. 
 The algebra   $ \chAn ( S^3 ) = Mat_N(C) $ should therefore be 
 distinguished from $ {\hat { \cal A} }^{(g)}_n  ( S^3 ) $,
  the sub-algebra generated by the coordinates.   
 The algebra   $ \hAg (S^3 )$ 
 contains the symmetric representations with unit multiplicity, 
 whereas $\chAn (S^3 )$ contains them with multiplicity $2$.     

  As in the case of the fuzzy 4-sphere, we need to 
 apply a projection if we want to recover the spectrum of 
 representations of the classical 3-sphere in the large 
 $n$ limit. Both $ \chAn ( S^3 )$ and $ \hAg (S^3 )$ 
 contain all the desired representations but contain extra
 representations as well.
 The algebra $ \cAn (S^3) $ can be defined as a projection 
 of the Matrix algebra, where we restrict $r_2 = 0 $, 
 and we require the matrices to be invariant under conjugation 
 by the permutation matrix which exchanges $ \cRp $ and $\cRm$, 
 and $P_+$ with $P_-$.

 \newsec{Higher Odd Spheres :  The five-sphere }
 
 For the fuzzy sphere $ S^{2k -1 }$, the coordinates 
 are matrices acting in a reducible representation 
 $ \cRp \oplus \cRm$ of $SO(2k)$. The weights of 
   $\cRp$ are $ \vec r = ( { n \over 2 },{ n \over 2 }, \cdots ,{ n \over 2 }, 
 { 1 \over 2 }  ) $ and those of  
  $ \cRm $ are $ \vec r = ( { n \over 2 },{ n \over 2 }, \cdots , { n \over 2
 }, { -1 \over 2} ) $, where $ \vec r $ is a
 $k$ dimensional vector. We discuss in detail the case of 
 the five sphere below. 

 Weights of $\cRp$ are 
$ \vec r  = ( { n \over 2 },{ n \over 2 }, { 1 \over 2 } ) $ 
and those 
 of $ \cRm $ are $ \vec r = ( { n \over 2 },{ n \over 2 }, { -1 \over 2
 } ) $.  From \dimse\ the dimension of each is
\eqn\dimssp{ 
 {  N \over 2 } \equiv Dim( \cRm) = Dim( \cRp ) 
   =  { 1 \over 192 } ( n+1 ) ( n+3 )^3  (n+5).  } 
Matrices acting on $\cR $ can be decomposed into four blocks 
$ End( \cRp ) $,  $ End( \cRm ) $, $Hom( \cRp, \cRm )$ 
 and    $Hom( \cRm, \cRp )$.  $ End( \cRm ) $
is related to $ End( \cRp ) $ by changing $P_+$ to $ P_-$. 
 Similarly   $Hom( \cRm, \cRp )$ is related to  $Hom( \cRp, \cRm )$. 
 
\subsec{ $End( \cRp )$ } 
 Consider operators of 
 the form 
\eqn\frmop{\eqalign{
 \sum_{\vec s } ~~~~~ &\rho_{s_1} ( \Gamma \Gamma \Gamma  P_+ ) \cdots
  \rho_{s_{p_1} } ( \Gamma \Gamma \Gamma  P_+ ) \cr 
  & \rho_{s_{p_1 + 1} } ( \Gamma \Gamma P_+ ) \cdots  \rho_{s_{p_1 +
 p_2^+ }}  ( \Gamma \Gamma P_+ ) \cr 
& \rho_{s_{p_1 + p_2^+ + 1  }}( \Gamma \Gamma P_-  ) 
 \cdots \rho_{s_{p_1 + p_2^+ + p_2^-  }}( \Gamma \Gamma P_-  ) \cr 
 &  \rho_{s_{p_1 + p_2^+ + p_2^- +1  }}( \Gamma  P_+  ) \cdots  
 \rho_{s_{p_1 + p_2^+ + p_2^- + p_3^+  }} ( \Gamma P_+ ) \cr 
 &  \rho_{s_{p_1 + p_2^+ + p_2^- + p_3^+ + 1   }} ( \Gamma P_- ) \cdots 
  \rho_{s_{p_1 + p_2^+ + p_2^- + p_3^+ +  p_3^-  }} ( \Gamma P_- )  \cr }}
 This gives a non-zero action  on vectors in $Sym(  V^{\otimes n } )  $ 
 which contain $p_1 + p_2^+ + p_3^+ + p_4^{+} = { (n+1)\over 2 } $ factors 
 of positive chirality, 
 where $p_4^+$ can be any integer between $0$ and $  { (n+1) \over 2 }
 - p_1 + p_2^+ + p_3^+ $. 
 The outgoing vector has $ p_2^+ + p_3^- + p_4^+ = (n+1)/2 $ 
 vectors of positive chirality. 
 These equations imply that $ p_3^- = p_3^+ + p_1$. 
 Writing $p_3^+ = k$, we see that 
\eqn\ptrmp{  p_3^- = k + p_1. }  
The expression of the form \frmop\ 
corresponds to a representation of weight given by  
$  \vec r = ( p_1 + p_2^+ + p_2^- + p_3^+ + p_3^- ,  p_1 + p_2^+ +
p_2^-, p_1 ) $. To be more explicit, we assign $ SO(6)$ indices 
 to the $ \Ga$ and we multiply by a traceless tensor with 
 the appropriate Young symmetry as in \frmgi. 

 We can obtain  the dimension $ Dim( { \vec r } )$ 
of the representation using \dimse.  
The summation with the appropriate range for the different indices 
is : 
\eqn\rng{ \sum_{p_1=0}^{ {n-1 \over 2 } } \sum_{ k=0}^{{n-1\over 2 } -  
p_1 } \sum_{p_2^{+}= 0 }^{ {( n+1)\over 2 } - k - p_1 } 
\sum_{p_2^-=0}^{ {(n-1)\over 2 } - k - p_1 } 
 Dim( \vec r ) } 
The upper bound on $p_1$ follows  because
$p_3^- \ge p_1 $ ( as is evident from \ptrmp\ ) 
and $p_3^- \le {(n-1)\over 2}$. 
The upper bound on $k$ is similarly due to the fact 
 that the operator \frmop\ hits 
 $k + p_1$ copies of $P_-$ and this is  bounded by 
 ${  n- 1  \over 2 } $ in $\cRp$. We have included, in the above sum, 
 the representations with $ p_1 =0$ which are not 
 self-dual. 
 
 Next we consider anti-self-dual  operators of the form 
 \eqn\frmopi{\eqalign{
 \sum_{ {\vec s} }~~~~ &\rho_{s_1} ( \Gamma \Gamma \Gamma  P_- ) \cdots
  \rho_{s_{p_1} } ( \Gamma \Gamma \Gamma  P_- ) \cr 
  &\rho_{s_{p_1 + 1} } ( \Gamma \Gamma P_+ ) \cdots  \rho_{s_{p_1 + p_2^+ }}
 ( \Gamma \Gamma P_+ ) \cr 
& \rho_{s_{p_1 + p_2^+ + 1  }}( \Gamma \Gamma P_-  ) 
 \cdots \rho_{s_{p_1 + p_2^+ + p_2^-  }}( \Gamma \Gamma P_-  ) \cr 
  & \rho_{s_{p_1 + p_2^+ + p_2^- + 1  }}( \Gamma  P_+  ) \cdots  
 \rho_{s_{p_1 + p_2^+ + p_2^- + p_3^+  }} ( \Gamma P_+ ) \cr 
&  \rho_{s_{p_1 + p_2^+ + p_2^- + p_3^+ +  1   }} ( \Gamma P_- ) \cdots 
  \rho_{s_{p_1 + p_2^+ + p_2^- + p_3^+ +  p_3^-  }} ( \Gamma P_- ) \cr } } 
These correspond to weights 
$ \vec r =  ( p_1 + p_2^+ + p_2^- + p_3^+ + p_3^- ,  p_1 + p_2^+ +
p_2^-, - p_1 ) $. 
For these we sum in the range 
 \eqn\rngi{ \sum_{p_1=1}^{ {n-1 \over 2 } } \sum_{ k=0}^{{(n-2p_1-1) \over
2 }  } \sum_{p_2^{+}= 0 }^{ {( n+1)\over 2 } - k - p_1 }
\sum_{p_2^-=0}^{ {(n-1)\over 2 } - k - p_1 } 
 Dim( \vec r ) }
Performing the two sums in \rng\ and 
\rngi\ ( using Maple), 
 we get $ {N^2 \over 4 }$, which is the dimension 
of $ End( \cRp)$.

\subsec{ $ Hom( \cRp, \cRm ) $ } 

 Using notation similar to the discussion 
 in the previous section, we now have 
 $ p_3^- = p_1 + p_3^+ - 1 $. 
 The summations of $ D(p_1,p_2^+, p_2^-,p_3^- ) $  are done with 
\eqn\smrnge{\eqalign{ 
&    1 \le p_1 \le { n-1 \over 2 } \cr 
            &    0 \le p_3^- \le { n-1 \over 2 } - p_1 \cr 
            &    0 \le p_2^+  \le { n- 1 \over 2 } - p_3^- - p_1 \cr 
            &   0 \le { n- 1 \over 2 } - p_3^- - p_1,  \cr }} 
with multiplicity two, corresponding to self-dual and anti-selfdual 
representations. The summation of $ D(p_1=0 ,p_2^+, p_2^-,p_3^- ) $
is done with the above constraints \smrnge. These are representations 
 with less than three rows. 
 Adding up the dimensions ( using Maple ) in this range 
 we get ${ N^2 \over 4 }$, the dimension of   $ Hom( \cRp, \cRm ) $.  

\subsec{ General remarks } 
The same remarks as for the three-sphere 
 apply here, as far as the need to distinguish the matrix algebra
 $ \chAn( S^5 ) \equiv Mat_N( C) $, the algebra generated 
 by the coordinates $ \hAg ( S^5) $, and the algebra of 
 fuzzy spherical harmonics,  $ \cAn ( S^5 )$,   obtained 
 by a projection of the Matrix algebra.

\newsec{ Projections of Matrix Algebras 
 to Fuzzy  Spherical Harmonics and Non-associativity. } 
 
 The Matrix Algebras associated to the various even and 
 odd fuzzy spheres $S^m$  have been decomposed, in previous sections, 
 in terms of representations of $SO(m+1)$. 
 The Matrix algebras contain too many representations 
 and have to be projected to recover the correct classical limit. 
 This was explained in section 4.3 in the context 
 of $S^4$. 
  It was already argued in \clt\ that in the large $n$ limit, 
  the algebra becomes commutative. Since an algebra can be commutative 
  but not associative we need to show that the algebra $\cAn(S^4)$
  is indeed associative in the large $n$ limit, to prove that 
  we correctly recover a classical sphere.

 Let us start with three elements $A,B,C$ of the Matrix algebra 
 in \sfsum,  
 which are all symmetric, i.e they are sums of 
 operators transforming in representations 
 with $r_2 =0$. This allows us to write 
\eqn\stsym{\eqalign{   P(A ) & = A \cr 
                     P(B ) & = B \cr
                     P(C ) & = C.  \cr}} 
 The question of associativity of the projected 
 product involves a comparison of $ ( (AB)_+ C )_+ $ 
 with $ (A (BC)_+)_+$. It is useful to recall 
 that 
\eqn\assocp{  ( (AB) C ) =  (A (BC) ) } 
By separating each side into symmetric representations 
 with $r_2 =0 $ and non-symmetric representations with 
$r_2 \ne 0$, we deduce 
\eqn\assocpi{  ( (AB) C )_+ =  (A (BC) )_+  }
and 
 \eqn\assocpii{  ( (AB) C )_- =  (A (BC) )_- }
Now consider the expression   $( (AB) C )_+$. We can write
$ ( (AB) C )_+ =  ( (AB)_+ C )_+ + ( (AB)_- C )_+$ where we have
 separated the symmetric and non-symmetric representations in $AB$. 
Similarly we can write 
$ ( A ( B C) )_+ =  ( A ( B C)_+ )_+ +  ( A ( B C)_- )_+$. 
Using these two expressions, along with \assocpi, we deduce that
 the failure of associativity is given by :  
\eqn\conas{    ( (AB)_+ C )_+ -  (A (BC)_+)_+ =  ( A ( B C)_- )_+ - 
 ( (AB)_- C )_+ } 
 This shows that the failure of associativity of the projected product 
 defined in \newpd\ is related to the fact 
 that the non-symmetric representations can mutiply 
 symmetric representations to give symmetric representations.
 Equivalently, the non-symmetric representations do not form an
 ideal. What will be significant in the following is that 
 both terms on the RHS of \conas\ contain elements ( $ (AB)_-$ and 
 $(BC)_-$ ) obtained by coupling two symmetric representations to 
 non-symmetric representations.

 Now a simple combinatoric argument can be used to 
 show that the non-associativity vanishes in the large $n$ 
 limit. This follows from the fact the couplings  $(AB)_-$ 
 are subleading.  Consider, 
 for concreteness, an operator among those in \frmg\  
 of the type $ \sum_{\vec s } \rho_{s_1} (  \Ga ) \rho_{s_2}  ( \Ga )
 $, associated with 
 the representation $\vec r = (2,0)$.
  When we square such 
 an operator we get, among other things, 
  the identity matrix with a  coefficient of order $n^4$.
 This comes the fact that the product contains 
 terms where the $ \Ga$ matrices are acting in four different 
 vector spaces in $ Sym^{\otimes n } ( V) $. For appropriate 
 choice of $SO(5) $ indices on the $ \Ga$'s this will be the identity 
 matrix with coefficient of order $1$. Now there is a 
 factor of ${ n \choose 4 } $ from the number of ways of choosing 
 four distinct factors in the tensor product. This grows like $n^4$. 
 We normalize the operators such that the coefficient of the identity 
 matrix is always $1$. The correctly normalized operator 
 goes like    $ { 1 \over n^2 } \sum_{\vec s }  \rho_{s_1} (  \Ga ) 
\rho_{s_2}  ( \Ga ) $
 in the large $n$ limit. In general the normalizing factor 
 for operators corresponding to Young diagrams with $r_1$ columns
 ( i.e a first row of length $r_1$ ),   
 obeying $ r_1 \ll n$ behaves like $ { 1 \over n^{r_1} }$.

 In the product of two copies of $ \sum_{\vec s }  
~~ \rho_{s_1} (  \Ga ) \rho_{s_2}  ( \Ga ) $, we will 
 have terms of the form 
  $ \sum_{\vec s } ~~~   \rho_{s_1} (  \Ga ) \rho_{s_2}  ( \Ga ) 
 \rho_{s_3} (  \Ga ) \rho_{s_4}  ( \Ga ) $. 
 These appear with coefficient of order $1$. 
 It follows that the product of two copies of the 
 correctly normalized operators 
 related to the Young Diagram with $ \vec r = (2,0)$ 
 contain the operators associated to the Young Diagram 
 with $ \vec r = (4,0)$ 
 with coefficient of order $1$. In  the product there 
 are also operators of the form 
\eqn\frmtl{ 
 \sum_{\vec s }  \rho_{s_1} (  \Ga ) \rho_{s_2}  ( \Ga^{\mu}   ) 
 \rho_{s_3} (  \Ga  ) \rho_{s_4}  ( \Ga^{\mu}  )  } 
 Now using the equation 
\eqn\contid{ \Gamma^{\mu} \otimes \Gamma^{\mu}  = 1 \otimes 1 } 
for operators in $ Sym( V ^{\otimes 2 } ) $, and summing over 
the $s_2,s_4$ indices we get an expression of 
 the form  
\eqn\frmtl{ 
 n^2 \sum_{\vec s }  \rho_{s_1} (  \Ga ) \rho_{s_3}  ( \Ga  ),   } 
where we have exhibited the  large $n$ behaviour of the coefficient. 
 Again after correctly normalizing we see that 
 the coupling of the representation $ \vec r = ( 2,0) $ to
 itself contains the symmetric representation $ \vec r = (2,0) $ 
 with coefficient of order 1. 
 
 This should be contrasted with the 
 coupling from pairs of symmetric representations 
 to nonsymmetric representations which is subleading in 
 ${1\over n} $. In the product considered above there are 
 terms where $s_1=s_4 $ which leads to 
 \eqn\frmtli{ 
 \sum_{\vec s }  \rho_{s_1} ( \Ga   \Ga ) \rho_{s_2}  ( \Ga  ) 
 \rho_{s_3} (  \Ga   )  }  
 If we keep the $SO(5)$ 
indices on the two $ \Ga$ in $\rho_{s_3} (  \Ga \Ga  )$, we see that  
 this includes 
 an operator transforming in the representation 
 $ \vec r = ( 3,1 )$. The coupling of the un-normalized operators 
 is of order $1$. After normalizing we have a factor of $ {1\over n^4} $ from 
 the normalizations of the factors in the product, 
 and a factor of $ { 1 \over n^3 }$ from the normalization of the
 factors in the result. Therefore the coupling of 
 a pair of normalized symmetric operators associated with 
 the Young diagram $ \vec r = (1,1)$ to the normalized 
 operator corresponding to $ \vec r = (2,1) $ is of order $ { 1 \over
 n }$
 in the large $n$ limit.     
 To conclude the correctly 
 normalized symmetric operators  couple to non-symmetric 
 operators with coefficients of subleading order in the 
 large $n$  expansion.  While we discussed a particular 
 example, it can easily be seen that this combinatorics is generic. 
Since \conas\ shows that the failure 
 of associativity can be expressed in terms of these 
 fusions of symmetric representations $ A,B$ into non-symmetric 
 representations, we have proved that the non-associativity 
 disappears in the leading  large $n$ limit. 
 It does, however persist, in the ${1 \over n} $ expansion.

 These observations were made above, for concreteness, 
 in the context of the fuzzy four-sphere, but 
 are clearly applicable to all the higher 
 spheres $ \cAn( S^m) $ with $ m > 2 $. 
 As we observed, in an earlier section, this appearance of 
 non-associativity is inevitable, if we want a non-commutative sphere 
 which has as its algebra of functions a finite set of symmetric 
 traceless representations. The sum of dimensions of these
 representations typically is not an absolute square, 
 and hence they cannot be realized as 
  a finite dimensional associative algebra with unit element
 over the complex numbers \gdwa.

 \newsec{ Remarks on Combinatorics And Geometry of Diverse fuzzy spheres } 

 While all the even and odd spheres 
 can be related in an $SO(m+1)$ covariant manner
 to matrix algebras, only for the two-sphere 
 the matrix algebra coincides with the algebra 
 of functions on the fuzzy sphere. For $m > 2$, 
 the Matrix algebra contains more degrees of freedom.  
 In this section we investigate  the extra 
 degrees of freedom,  from the point of view of physics 
 on the sphere. 

 To this end, we first summarize some facts. 
 For $ \cAn( S^{2k} )$, 
  $ R^2 \equiv \sum_{\mu} ( G_{\mu} G_{\mu } ) = n^2 + 2kn$. 
 For    $ \cAn ( S^{2k-1 } )$, 
 $ R^2 \equiv \sum_i ( \hat G_{i } \hat G_{i  } ) 
= {1\over 2 } ( n+1) ( n + 2k -1)  $
\foot{ In the first version of this paper this was given 
as $n^2 + 2kn -1$, which  is actually the eigenvalue of 
 a related quantity $\sum_{r,s} 
 \sum_{i}  \Pr \rho_r (\Ga_i )  \rho_s (\Ga_i )
 \Pr $ which should be distinguished from
 $   \sum_i ( \hat G_{i } \hat G_{i  } ) = 
\sum_{r,s} \sum_{i}  \Pr \rho_r (\Ga_i ) \Pr  \rho_s ( \Ga_i ) \Pr  $.
They are both operators within $ Sym (  V^{\otimes n }  ) $ but the
 quantity which should be correctly called the radius is the 
 latter which acts entirely within $\cR$.  }.
 In all cases  $ R \sim n$ at large $n$.  
 For even fuzzy spheres, 
the size of the matrix algebra 
 $ \chAn ( S^{2k} )$ grows like   $n^{{ (k) ( k+ 1 ) }}$. 
 For odd fuzzy spheres, the size of $ \chAn ( S^{2k-1 } )$ grows like 
  $n^ { { ( k-1 ) ( k+2 )  }} $. 
 Note that $S^1$ is a special case so the above 
 formulae should not be applied to $S^1$. One may attempt 
 to define a non-commutative $S^1$ using the techniques 
 of this paper, but such a non-commutative $S^1$ admits 
 no classical limit which relies on large irreducible 
 reprsentations of the $SO$ groups for the higher dimensions.

 The combinatorics of the 
fuzzy 2-sphere is fairly intuitive. 
 The radius grows like $n$, and the number of degrees 
 of freedom like $n^2$. In physical applications such as 
 polarization, $ R \sim L n $, where $L$ is a physical length 
 scale, which can be, for example, the string scale or the eleven
dimensional 
 Planck length, depending 
on the context. Using semiclassical intuition,  a particle 
moving on such a 
 sphere  would have a configuration space with volume growing  like
$n^2 L^2 $. This agrees with the number of degrees of freedom 
 of the Matrix algebra. So the Matrix algebra describes 
 fairly conventional physics on the sphere. 

 Consider now the fuzzy 4-sphere. 
  The radius of the fuzzy sphere behaves like 
  $R \sim n $. The volume behaves like 
  $ V \sim n^m$ for $S^m$. And this is indeed 
 the behaviour of the projected algebra 
 $ \cAn ( S^4 )$. 
 The Matrix algebra $ \chAn ( S^4 )$  has $n^6$ degrees of freedom. 
 This looks like it is too numerous for a conventional explanation  
 in terms of particles on the sphere. It is useful to recall 
 that in physical applications in matrix theory \clt\ 
 or in descriptions of fuzzy funnels \fun, $n$ is the number of 5-branes and $n^3$ is the number 
 of instantons. In many instanton moduli space problems \insmod, 
 instantons of $U(n)$ behave as if they can fractionate 
 into particles of instanton number $ { 1\over n } $. 
 Fractional instantons also show up in contexts like 
 \mal, \bdie, \zk , \buk. 
 We might expect here to understand the degrees 
 of freedom in terms of $n^4$ fractional instantons moving 
 on an $S^4$ of radius $n$.  
 Interestingly, the volume of the configuration space 
 is now too big,  i.e grows like $n^8$. If each fractional instanton 
 were associated with $n^2$ degrees of freedom, we would 
 have the correct counting. This is suggestive of a 
 picture of the low-energy dynamics where each fractional instanton 
 has a two-dimensional configuration space inside the four-sphere. 
 Precisely such a behaviour, where particles behave
 like two-dimensional  extended objects with a 
 two-dimensional transverse space,   was found for giant gravitons 
 on $S^4$ in \mst.  It would be very interesting to explore 
 possible connections with \mst\ and the fuzzy four-sphere 
 defined using the projection,  in more detail. 
 Quantum mechanics on the fuzzy four-sphere, generalizing 
 \napo,  and more detailed information about 
 instantons on $S^4$, would be a probably  useful  starting point. 
  
 It is interesting to go back to the fuzzy 
 $S^2$ case in the light of the above discussion. 
 There, the physical application has $ N \sim n^2 $ zero-branes and a single 
 spherical two-brane \kata\myers. 
 Now moduli space studies of magnetic flux suggest 
 fractionation is related to the greatest common denominator 
 of the rank and the flux \ztor\konsch. In this case, we would then 
 be lead to expect degrees of freedom associated with a 
 single particle on the sphere. This indeed agrees with 
 the $n^2$   degrees of freedom of the Matrix algebra.

\newsec{ Summary and outlook }

 We developed in detail the relation between 
 higher dimensional fuzzy spheres $ \cAn( S^m ) $ 
 and Matrix algebras  $ \chAn ( S^m ) = End ( \cR ) $. 
 $ End( \cR )$ is the algebra of Matrices acting 
 on a representation of the isometry group $SO(m+1)$. 
 For $m $ even, $\cR $ is irreducible. For $m$ odd 
 $ \cR = \cRp \oplus \cRm $.  In both cases the matrix algebra 
 was decomposed into representations of 
 $ SO(m+1) $. Among these matrices is the algebra $ \cAn ( S^m )$ 
 which contains symmetric traceless reprentations with unit
 multiplicity. The product 
 structure on this algebra is obtained by using the matrix product
 followed by a projection. We showed that this product is
 non-associative ( section 4.3), but that the non-associativity 
 vanishes in the large $N$ limit ( section 9). 
  In the case of $m$ odd, it is also necessary to
 distinguish the algebra $ \hAg( S^{2k -1  } ) $ which is generated 
 by the coordinates $ {\hat G}^{i}$. This algebra is larger than 
 the algebra of spherical harmonics $\cAn( S^{2k -1 }  ) $ but smaller 
 than the matrix algebra $ \chAn ( S^{2k -1 } ) $. 

 The matrices $G$ are used to describe classical solutions 
 of brane actions. 
 Fluctuations can be an arbitrary $N \times N$ matrix. 
 An $SO(m+1) $ covariant description of the fluctuations requires the above 
 decomposition of the Matrix algebras. 
  
 The family of non-commutative spheres we described 
 admit embeddings 
\eqn\embd{ \cdots \rightarrow \chAn( S^{2k-1 } ) \rightarrow  \chAn
 ( S^{ 2k } )  \rightarrow  \chAn( S^{2k+1 } )\rightarrow \cdots } 
 and correspondingly 
\eqn\embds{ \cdots \rightarrow \cAn( S^{2k-1 } ) \rightarrow  \cAn
 ( S^{ 2k } ) \rightarrow  \cAn( S^{2k+1 } ) \cdots   }
 These embeddings follow from reduction of representations 
 of $SO(m+1)$ into representations of $SO(m)$. It will be interesting 
 to relate the details of these embeddings to the physical
 applications of the fuzzy spheres. 
 For example, the embeddings of fuzzy three-sphere 
 into four-sphere, may be expected \disc\ to be useful in studying 
 the connection between the polarization of D-instantons 
 in background five-form field strengths in type IIB \gomez\ 
 and the polarization of unstable $DO$-branes \zr. 
 It will also be interesting to study implications of these embeddings 
 in applications to Matrix theory \clt\ho\ and 
 to the ADS/CFT correspondence \antram\hrt\holi\bv\jmr.  
 
 While $ \cAn(S^m)$ is obtained from  $ \chAn ( S^m )$
 by a projection, it is of interest to understand the Matrix algebra 
 from the geometry of the noncommutative sphere.
 As a small  step in this direction, we considered the 
 case of $S^4$ and found hints ( section 10 ) that the degrees of freedom 
 of the Matrix algebra could be understood from a picture 
 where fractional instantons behave as extended objects, in a manner 
 reminiscent of giant gravitons \mst. The connection can be 
 explored further by developing quantum mechanics on these fuzzy
 spheres. 

 There have been a number of recent discussions of non-associative 
 algebras emerging from considering background NS-sector $H$-fields in string
 theory \corn\pmnass. Since spherical branes appearing in polarization
 effects 
 or Matrix theory, which motivated 
 this work, are a natural setting for non-constant field strengths
 ( RR field strengths in the more conventional applications ), it is
 natural to expect that there will be connections between the structure 
 of non-associativity present here and the one discussed in the above
 works. It will be interesting to see if a structure similar to 
 the one which came up here, involving non-associativity as 
 consequence of projection from an associative algebra, shows up 
 in the world-sheet  construction of exact  backgrounds describing 
 the spherical branes.  One similarity between 
 the discussion of \pmnass\ and the one here, 
 is that the bigger algebra involved here contains derivatives as 
 well as coordinates. For example the generators 
 of the $SO(m+1)$ isometries, 
 $G^{\mu \nu }$, are naturally thought as derivatives, but 
 belong to representations which are projected out in defining 
 $\cAn(S^m)$. 
 Non-associativity was also discussed 
 in the context of membrane quantization in \almy, 
  in string field theory in \horstro, in  the context of 
 $q$-spheres in \gms, and in superspace in \kpt.

  In \holi\ the fuzzy two-sphere was used to discuss 
 the entropy of black holes in four dimensions. 
 It may be fruitful to explore applications of 
 the higher spheres as developed here to similar questions. 
 In the case of the fuzzy four-sphere, some aspects 
 of the connection between the MAtrix algebra  and 
 spherical 4-brannes was understood using spherically symmetric 
 instantons. Similar constructions should be investigated  for the other 
 spheres.   
  Relations between these fuzzy spheres and those 
  based on quantum groups may exist along simlar lines 
 to the work in  \hidclass\ which related fuzzy and quantum group 
 symmetric spheres in two dimensions.  
  Another direction is to explore the detailed relation of these
  non-commutative spheres we have studied, inspired by  
  brane polarization problems and Kaluza-Klein truncation in 
 ADS/CFT, to the approach of \fgr, 
  where a deformation of phase space rather than a 
  deformation of configuration space ( closer to the  approach  taken
 here) is  studied.

\bigskip

\noindent{\bf Acknowledgements:}
 I would like to  thank for pleasant discussions S. Corley, V. Fock, 
 Z. Guralnik,  P. M. Ho,  A. Jevicki, V. Kazakov, D. Lowe and  J. Troost.
 This research was supported  by DOE grant  
 DE-FG02/19ER40688-(Task A).

\newsec{ Appendix } 

 In associating operators to representations 
 of $SO(4)$ in our discussion of the $3$-sphere, 
 we used operators of the form 
\eqn\apo{ 
 \rho_{s_1} ( \Gamma \Gamma P_+ ) \rho_{s_2}  ( \Gamma \Gamma P_+  ) 
    \cdots \rho_{s_{p_1} } ( \Gamma \Gamma P_+ ) \rho_{s_{p_1+1} }
  ( \Gamma ) \rho_{s_{p_1+2 } }  ( \Gamma ) \cdots \rho_{s_{p_1+p_2}}
 ( \Gamma ) }  
 For simplicity we have written out indices 
 on the $\Gamma$ matrices, but it is understood that 
 the full expression adds indices, applies the Young symmetrizer
 and subtracts a trace as appropriate to the Young diagram appropriate 
 to the above pattern of $\Gamma$. 
 We observed that, without loss of generality, we can add 
 a number of $P_+$ and an equal  number of $P_-$ when we are 
 considering $End(R_+)$. We could also have all $P_-$ multiplying 
 the first set of $ \Gamma \Gamma $ in \apo\ and an equal number 
 of $P_+$ and $P_-$ attached to the second set of $ \Gamma$ in \apo.  

 We never needed to consider operators of the form 
\eqn\apt{ 
 \rho_{s_1} ( \Gamma \Gamma P_+ ) \rho_{s_2}  ( \Gamma \Gamma P_-  ) 
    \cdots \rho_{s_{p_1} } ( \Gamma \Gamma P_+ ) \rho_{s_{p_1+1} }
  ( \Gamma ) \rho_{s_{p_1+2 } }  ( \Gamma ) \cdots \rho_{s_{p_1+p_2}}
 ( \Gamma ) }
 where some of the set of $ \Gamma \Gamma $ are multiplying 
 $P_+$ and some are multiplying $P_-$. The reason is that 
 these are expected to vanish identically. The quickest way to 
 see this is that operators with all $P_+$ or all $P_-$ 
 are naturally associated with Young Diagrams of 
 $SO(4)$ with the second row of length $p_1 $ or
 $-p_1$. There is no obvious way to associate a representation 
 to the operators involving mixing of $ P_+ $ and $P_-$ 
 as in \apt. The same rule applies to  
 expressions of the form $ \Gamma \Gamma \Gamma $ in $ SO(6)$, 
 and expressions of the form $ \Gamma \Gamma \Gamma \Gamma $ in
$SO(8)$, but expressions involving $ \Gamma \Gamma $ in the higher 
 rank $SO$ are not restricted to multiplying projectors 
 os a single parity. 

 We will these points explicitly in some simple examples. 
 We work with $SO(4)$. The representation with 
 $ {\vec r }= ( 1,1 )$ corresponds to the operators
 $$ \sum_{s} \rho_{s_1} ( \Gamma \Gamma P_+ ) $$ 
 This representation has dimension $ 3 $, and 
 indeed there are three self-dual combinations 
 we can write 
\eqn\apex{\eqalign{ 
 \sum_{s} \rho_{s_1} ( \Gamma^1 \Gamma^2 P_+ )
 &= \sum_{s} \rho_{s_1} ( (\Gamma^1 \Gamma^2  + \Gamma^3 \Gamma^4)P_+)
\cr 
 \sum_{s} \rho_{s_1} ( \Gamma^1 \Gamma^3 P_+ )
 &= \sum_{s} \rho_{s_1} ( (\Gamma^1 \Gamma^3  + \Gamma^4
\Gamma^2)P_+)\cr
& \sum_{s} \rho_{s_1} ( \Gamma^1 \Gamma^4 P_+ )
= \sum_{s} \rho_{s_1} ( (\Gamma^1 \Gamma^4  + \Gamma^2
\Gamma^3)P_+) \cr }} 
The representation $ \vec r = (1,-1) $ corresponds 
 to the operators  $$ \sum_{s} \rho_{s_1} ( \Gamma \Gamma P_- ) $$.
Again we have $3$ in agreement with the dimension : 
\eqn\apexi{\eqalign{ 
 \sum_{s} \rho_{s_1} ( \Gamma^1 \Gamma^2 P_- )
 &= \sum_{s} \rho_{s_1} ( (\Gamma^1 \Gamma^2  - \Gamma^3 \Gamma^4)P_-)
\cr 
 \sum_{s} \rho_{s_1} ( \Gamma^1 \Gamma^3 P_- )
 &= \sum_{s} \rho_{s_1} ( (\Gamma^1 \Gamma^3  - \Gamma^4
\Gamma^2)P_-)\cr
\sum_{s} \rho_{s_1} ( \Gamma^1 \Gamma^4 P_- )
 &= \sum_{s} \rho_{s_1} ( (\Gamma^1 \Gamma^4  + \Gamma^2 \Gamma^3)P_-) \cr }} 
 
Now consider $  \sum_{s} \rho_{s_1} ( \Gamma \Gamma P_+ )
\rho_{s_2} ( \Gamma \Gamma  P_+ )$. 
 These correspond to the irreducible rep. 
 with $ \vec r = ( 2,2 )$ which has dimension $5$. 
 So we should be able to exhibit $5$ independent 
 operators of the above form. We need to consider operators of the
form  $\sum_{s} \rho_{s_1} ( \Gamma^1 \Gamma^i P_+ )
 \rho_{s_2} ( \Gamma^1 \Gamma^j  P_+ )$ with $ 2 \le i \le j \le 4 $. 
 There are a total of six such operators. We have to show that 
 there is one relation. 
 Take operators of the form   $\sum_{s} \rho_{s_1} ( \Gamma^1 \Gamma^2 P_+ )
 \rho_{s_2} ( \Gamma^1 \Gamma^2  P_+ )$. 
 To associate to an irrep of $SO(4)$ we must do a 
 symmetrization procedure and a subraction of the trace 
 appropriate to the corresponding Young diagram. 
 The vertical  antisymmetry is automatically present given the 
 properties of $ \Gamma $ matrices. The horizontal symmetry 
 is guaranteed by the sums over $s_1$ and $s_2$. 
 Let us consider the tracelessness condition  which gives : 
\eqn\trcon{\eqalign{& 
  \sum_{s} \rho_{s_1} ( \Gamma^1 \Gamma^2 P_+ )~
 \rho_{s_2} ( \Gamma^1 \Gamma^2  P_+ ) ~~
 + ~~ \rho_{s_1} ( \Gamma^2 \Gamma^2 P_+ )
 \rho_{s_2} ( \Gamma^2 \Gamma^2  P_+ ) \cr
& + ~~ \rho_{s_1} ( \Gamma^3 \Gamma^2 P_+ )~
 \rho_{s_2} ( \Gamma^3 \Gamma^2  P_+ )
 ~~+ ~~ \rho_{s_1} ( \Gamma^4 \Gamma^2 P_+ )~
 \rho_{s_2} ( \Gamma^4 \Gamma^2  P_+ ) = 0 \cr }} 
 The second term belongs to an irreducible rep. 
 with fewer boxes, so we can drop it in considering 
 relations involving $ \vec r = (2,2)$. Alternatively we 
 can use antisymmetry along the columns of the Young diagram to drop
 such a term. So the relation we get is 
  Let us consider the tracelessness condition  which gives : 
\eqn\trconi{\eqalign{& 
  \sum_{s} \rho_{s_1} ( \Gamma^1 \Gamma^2 P_+ )
~ \rho_{s_2} ( \Gamma^1 \Gamma^2  P_+ )~~
 + ~~  \rho_{s_1} ( \Gamma^1 \Gamma^4 P_+ )
 \rho_{s_2} ( \Gamma^1 \Gamma^4  P_+ ) \cr 
&  + ~~  \rho_{s_1} ( \Gamma^1 \Gamma^3 P_+ ) ~
 \rho_{s_2} ( \Gamma^1 \Gamma^3  P_+ ) = 0 \cr   }} 
Note that if we had started with 
$\sum_{s} \rho_{s_1} ( \Gamma^1 \Gamma^3 P_+ )
 \rho_{s_2} ( \Gamma^1 \Gamma^3  P_+ ) $
and tried to impose the traceless condition we would get the same
 equation. So we have one relation and six operators leaving us 
 with $5$ independent ones. 

On the other hand consider an operator of the form 
$  \sum_{s} \rho_{s_1} ( \Gamma^1 \Gamma^2 P_+ )
 \rho_{s_2} ( \Gamma^1 \Gamma^2  P_- ) $. Applying a tracelessness
 condition we get : 
\eqn\tlesi{\eqalign{
 & \sum_{s} \rho_{s_1} ( \Gamma^1 \Gamma^2 P_+ )
  \rho_{s_2} ( \Gamma^1 \Gamma^2  P_- ) 
   + \rho_{s_1} ( \Gamma^3 \Gamma^2 P_+ )
  \rho_{s_2} ( \Gamma^3 \Gamma^2  P_- ) 
  + \rho_{s_1} ( \Gamma^4 \Gamma^2 P_+ )
  \rho_{s_2} ( \Gamma^4 \Gamma^2  P_- )  \cr 
& =  
 \sum_{s} \rho_{s_1} ( \Gamma^1 \Gamma^2 P_+ )
  \rho_{s_2} ( \Gamma^1 \Gamma^2  P_- ) 
   - \rho_{s_1} ( \Gamma^1 \Gamma^4 P_+ )
  \rho_{s_2} ( \Gamma^1 \Gamma^4 P_- ) 
  - \rho_{s_1} ( \Gamma^1 \Gamma^3 P_+ )
  \rho_{s_2} ( \Gamma^1 \Gamma^3  P_- )  = 0   \cr }}
If we start with $  \sum_{s} \rho_{s_1} ( \Gamma^1 \Gamma^3 P_+ )
 \rho_{s_2} ( \Gamma^1 \Gamma^3  P_- ) $, we get a similar equation
 with different signs : 
\eqn\tsii{  
 \sum_{s} - \rho_{s_1} ( \Gamma^1 \Gamma^2 P_+ )
  \rho_{s_2} ( \Gamma^1 \Gamma^2  P_- ) 
   - \rho_{s_1} ( \Gamma^1 \Gamma^4 P_+ )
  \rho_{s_2} ( \Gamma^1 \Gamma^4 P_- ) 
   + \rho_{s_1} ( \Gamma^1 \Gamma^3 P_+ )
  \rho_{s_2} ( \Gamma^1 \Gamma^3  P_- )  = 0    }
And finally starting from $  \sum_{s} \rho_{s_1} ( \Gamma^1 \Gamma^3 P_+ )
 \rho_{s_2} ( \Gamma^1 \Gamma^3  P_- ) $ we get 
 \eqn\tlesi{  
 \sum_{s} - \rho_{s_1} ( \Gamma^1 \Gamma^2 P_+ )
  \rho_{s_2} ( \Gamma^1 \Gamma^2  P_- ) 
   + \rho_{s_1} ( \Gamma^1 \Gamma^4 P_+ )
  \rho_{s_2} ( \Gamma^1 \Gamma^4 P_- ) 
   -\rho_{s_1} ( \Gamma^1 \Gamma^3 P_+ )
  \rho_{s_2} ( \Gamma^1 \Gamma^3  P_- )  = 0    }
These three independent conditions ensure the vanishing 
 of all operators of the form 
 $  \sum_{s} \rho_{s_1} ( \Gamma^1 \Gamma^i P_+ )
 \rho_{s_2} ( \Gamma^1 \Gamma^i  P_- ) $, 
 satisfying the tracelessness and symmetry properties 
 of the candidate corresponding Young diagram. 
For operators of the form 
$  \sum_{s} \rho_{s_1} ( \Gamma^1 \Gamma^i P_+ )
 \rho_{s_2} ( \Gamma^1 \Gamma^j  P_- ) $ 
with $i \ne j$ the symmetry in $i$ and $j$ 
 combined with tracelessness again set the operator 
 to zero. 

 We have discussed the case of $SO(4)$ in detail, 
 but we expect similar arguments to work for 
 higher rank even $SO(2k)$ groups ruling out
 operators involving both $P_+ $ and $P_-$ 
 attached to a string of $k$ $\Gamma$ matrices.

\bigskip

\listrefs
\end